\documentclass{aa}  

\usepackage{graphicx}
\usepackage{txfonts}
\usepackage[]{hyperref}
\usepackage{xcolor}
\usepackage{soul}
\usepackage{booktabs}
\usepackage{tabularx}

\newcommand{\Rer}{\ensuremath{R_{\oplus}}}
\newcommand{\Mer}{\ensuremath{M_{\oplus}}}
\newcommand{\Msun}{\ensuremath{M_{\odot}}}
\newcommand{\Rpl}{\ensuremath{R_{\rm pl}}}
\newcommand{\Mpl}{\ensuremath{M_{\rm pl}}}
\newcommand{\Teq}{\ensuremath{T_{\rm eq}}}

\begin{document} 

   \title{Precise photoionisation treatment and hydrodynamic effects in atmospheric modelling of warm and hot Neptunes}
   \subtitle{}
   \titlerunning{Photoionisation and hydrodynamic effects in atmospheres of warm and hot Neptunes}
   \author{
   Daria Kubyshkina\inst{1} $^{\href{https://orcid.org/0000-0001-9137-9818}{\includegraphics[scale=0.5]{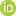}}}$,
   Luca Fossati\inst{1} $^{\href{https://orcid.org/0000-0003-4426-9530}{\includegraphics[scale=0.5]{FIGURES/orcid.jpg}}}$, 
   Nikolai V. Erkaev\inst{2} $^{\href{https://orcid.org/0000-0001-8993-6400}{\includegraphics[scale=0.5]{FIGURES/orcid.jpg}}}$
   }
   
   \authorrunning{Kubyshkina et al.}
   
   \institute{
        Space Research Institute, Austrian Academy of Sciences, Schmiedlstrasse 6, A-8042 Graz, Austria\\
        \email{daria.kubyshkina@oeaw.ac.at} 
   \and
        Institute of Computational Modelling of the Siberian Branch of the Russian Academy of Sciences, Krasnoyarsk, Russian Federation
   }

   \date{Received XXX; accepted XXX}

  \abstract
   {Observational breakthroughs in the exoplanet field of the last decade motivated the development of numerous theoretical models, such as those describing atmospheres and mass loss, which is believed to be one of the main drivers of planetary evolution.}%
   {We aim to outline for which types of close-in planets in the Neptune-mass range the accurate treatment of photoionisation effects is most relevant, particularly with respect to atmospheric mass loss and the parameters relevant for interpreting observations, such as temperature and ion fraction.}
   {We developed the CHAIN (Cloudy e Hydro Ancora INsieme) model, which combines our 1D hydrodynamic upper atmosphere model with the non-local thermodynamical equilibrium (LTE) photoionisation and radiative transfer code Cloudy accounting for ionisation, dissociation, detailed atomic level populations, and chemical reactions for all chemical elements up to zinc. The hydrodynamic code is responsible for describing the outflow, while Cloudy solves the photoionisation and heating of planetary atmospheres. We apply CHAIN to model the upper atmospheres of a range of Neptune-like planets with masses between 1 and 50 M$_{\oplus}$, varying also the orbital parameters.}
   {For the majority of warm and hot Neptunes, we find slower and denser outflows, with lower ion fractions, compared to the predictions of the hydrodynamic model alone. Furthermore, we find significantly different temperature profiles between CHAIN and the hydrodynamic model alone, though the peak values are similar for similar atmospheric compositions. The mass-loss rates predicted by CHAIN are higher for hot, strongly irradiated planets and lower for more moderate planets. All differences between the two models are strongly correlated with the amount of high-energy irradiation. Finally, we find that the hydrodynamic effects impact significantly ionisation and heating.}
   {The impact of the precise photoionisation treatment provided by Cloudy strongly depends on the system parameters. This suggests that some of the simplifications typically employed in hydrodynamic modelling might lead to systematic errors when studying planetary atmospheres, even at a population-wide level.}

   \keywords{
   planets and satellites: atmospheres -- 
   hydrodynamics -- 
   radiation mechanisms: general
   }

   \maketitle
   \nolinenumbers
%
\section{Introduction}\label{sec::intro}
There is strong theoretical and observational evidence demonstrating that atmospheric escape takes a significant role in shaping the population of intermediate-mass planets \citep[see e.g.][]{owen_wu2013,owen_wu2017,fulton2017,ginzburg2018,jin2018,gupta_schlichting2019,gupta_schlichting2020,sandoval2021,kubyshkina2022MR,ho_van_eylen2023} and in modifying their atmospheric composition \citep[see e.g.][]{Zahnle_kasting1986,hunten1987,albarede2007,odert2018,lammer2020}. This led to a substantial modelling effort aiming at explaining the structure and dynamics of planetary upper atmospheres and the mechanisms driving the escape \citep[see e.g. the reviews of][and references therein]{gronoff2020,owen2020review}. This effort is being supported by multi-wavelength transmission spectroscopy observations that help to guide and constrain the theoretical models \citep[e.g.][]{vidal-madjar2003,fossati2010,haswell2012,ehrenreich2015,Lavie2017,Allart2018,Spake2018,Nortmann2018,carleo2020,Vissapragada2020,cubillos2020,Orell-Miquel2022}. 

The assumptions taken when computing the thermal and chemical structure of planetary atmospheres play an important role in the interpretation of the observations. For example, uncertainties in the atmospheric ion fraction, caused e.g. by inaccuracies in the stellar spectral energy distribution or in the theoretical model, can lead to significant uncertainties in the mass-loss rates \citep[e.g.][]{guo_benjaffel2016,odert2018,kubyshkina2022_3d} or in the temperature of the outflow \citep{Vissapragada2022_constr_p-wind} derived from observations. Another physical effect that in upper atmosphere models needs to be taken into account, particularly when comparing with observations, is the interaction with the stellar wind \citep[e.g.,][]{Bisikalo2013,Schneiter2016,Carroll-Nellenback2017,erkaev2017sw,Villarreal2018,Esquivel2019,Khodachenko2019,Khodachenko2021,Villarreal2021,carolan2021,kubyshkina2022_3d,Cohen2022}. 

While the existing hydrodynamic upper atmosphere models treat similarly well the transport of particles in the atmosphere, there are significant differences among models in the treatment of energy balance (i.e. heating and cooling). The latter is particularly challenging to model given the large number and diversity of non-local thermodynamical equilibrium (NLTE) processes that are involved.

In this study, we present a new code called CHAIN (Cloudy e Hydro Ancora INsieme) for modelling planetary upper atmospheres, which combines the hydrodynamic model of the atmospheric outflow \citep{kubyshkina2018grid} with the hydrostatic NLTE radiative transfer solver Cloudy \citep{ferland2017}. This approach was previously used by \citet{salz2015,salz2016} and allows one to significantly increase the accuracy of the photochemistry in the hydrodynamic simulations at relatively low computational costs while keeping the atmospheric temperature-pressure structure consistent. {To test, how the model performs under different conditions, we} then apply the new code to compute a small grid of model planets, ranging from terrestrial-like to sub-Saturn-like, further changing their equilibrium temperatures and level of high-energy stellar irradiation (X-ray and extreme ultraviolet, EUV; together referred to as XUV), to finally compare the results to the predictions of the sole hydrodynamic model driven by hydrogen chemistry. In this way, we aim at estimating in which regions of the parameter space the improved photochemistry is most relevant. %

This paper is organised as follows. In Section\,\ref{sec::model}, we describe the most relevant properties and features of the photoionisation solver Cloudy (\ref{sec::model_cloudy}) and of the hydrodynamic model (\ref{sec::model_hydro}) used to compile the new code. The details of the numerical implementation are given in Section\,\ref{sec::code}, including a discussion of the basic modelling scheme (\ref{sec::code_base}), the spectra used to represent the stellar irradiation (\ref{sec::code_SED}), the boundary conditions assumed in the code (\ref{sec::code_bord}), the considered atmospheric compositions (\ref{sec::code_muhe}), and the treatment of the hydrodynamic flow in the considered chemistry framework (\ref{sec::code_wind}). In Section \ref{sec::salz}, we validate our code by comparing it to the earlier results of \citet{salz2016a} and discuss some basic features of the model. In Section\,\ref{sec::gridtests}, we apply our model to the grid of model planets and discuss the results. We start by describing the set of planets in Section\,\ref{sec::gridtests_planets}, followed by a detailed discussion of the results obtained for one specific Neptune-like planet as an example (\ref{sec::gridtests_HE}). We further compare the predictions of the model throughout the entire grid assuming pure hydrogen atmospheres (\ref{sec::gridtests_grid}) and then discuss the effect of different atmospheric compositions (\ref{sec::gridtests_muhe}). Finally, we discuss the influence of the specific input stellar spectra on the results in Section\,\ref{sec::disscussion_spectra} and gather the conclusions in Section\,\ref{sec::conclusions}.
\section{Physical model}\label{sec::model}
\subsection{Photoionisation solver Cloudy}\label{sec::model_cloudy}

The Cloudy code is a photoionisation and spectral synthesis solver dedicated to modelling astrophysical environments in a wide range of temperatures and densities, such as gas clouds {or protoplanetary discs \citep{ferland1998,ferland2013,ferland2017}. The code has also been employed to model planetary upper atmospheres \citep[e.g. ][]{salz2016,salz2016a,linssen2022,fossati2021}, mainly focused on but not limited to hot Jupiters.} Within the code, the physical state of matter can range from bare nuclei to molecules and grains. The code solves the chemical, ionisation, and thermal structure of the gas irradiated by an external source (e.g. the host star for protoplanetary discs and planetary atmospheres) in a static density structure provided by the user and on the basis of this solution predicts, among other things, the physical properties of the gas and the transmitted spectrum.

From the numerical side, Cloudy is a one-dimensional (1D) NLTE spectral synthesis code. Therefore, the local equilibrium state is solved in the elementary volumes represented by the cells of the adaptive grid, which is generally set up on the basis of the gas density gradient. For these calculations, Cloudy accounts for ionisation and recombination processes, chemical reactions, and transitions between the energy levels of the atmospheric species. {The full collisional radiative models are applied at densities above $\sim10^3$\,${\rm cm^{-3}}$ \citep{ferland2017}, which condition is typically satisfied in the upper atmospheres of close-in sub-Neptune-like planets. }%

As a minimal input from the user, Cloudy needs the density structure and thickness of the considered gas cloud, its position relative to the {external radiation source, and the spectral energy distribution (SED) of the latter}. The default geometry of the simulation assumes the irradiation source to lie in the centre of the cloud (so-called spherical geometry), however, if the distance between the irradiation source and the cloud is much larger than the thickness of the cloud (typically the case for planetary atmospheres), the geometry changes to plane-parallel. 

{For (quasi-)static structures, Cloudy can compute} internally and self-consistently the thermal structure of the cloud, but it considers only the local thermal motion. Therefore, in the case of actively expanding planetary atmospheres, the thermal structure should be specified externally, as it is strongly dependent on adiabatic effects. Additionally, Cloudy enables the user to specify a velocity field (i.e. microturbulence or wind) in specific ways (see details in Section\,\ref{sec::code_wind}).

The stellar irradiation is set by the shape of the SED, which can be set by the user or adopted from an extensive internal library, and the intensity of the radiation, which is specified separately. The two define a unique radiation field, and up to 100 independent fields can be included in one simulation. 
When processing the incoming radiation, Cloudy considers the continuum and the line radiation as two separate heating sources. To solve the radiative transfer relative to the continuum, the code calculates the flux absorption within the gas accounting for the implemented opacity sources and scattering mechanisms. These, by default, include inverse bremsstrahlung, H$^-$ absorption, pair production, electron and Rayleigh scattering, photoabsorption by molecules, grain opacity (when included), and the photoelectric absorption by ground and excited states of the atomic species included in the model. Within the XUV wavelength range, which represents the main heating source for planetary upper atmospheres, the opacity is dominated by the ionisation of hydrogen species. Finally, the radiative transfer events are decoupled from the local equilibrium state and solved considering the escape probability mechanism \citep{castor1970,Elitzur1982}.

For planetary upper atmospheres, the main heating source is given by the photoionisation of the neutral atmospheric species, {which Cloudy defines as the energy input by the freed photoelectrons (given by the difference between the ionisation potential of the atom and the energy of the photon). It also accounts for secondary (collisional) photoionisation, including the ionisation from excited levels. The largest input (besides the ground state) is typically expected from the metastable 2S level.} The cooling processes include line radiation, bremsstrahlung cooling, and a wide range of chemical reactions. 
We consider heating and cooling in more detail when analysing our results in Section\,\ref{sec::gridtests_HE}.

Cloudy accounts for the 30 lightest chemical elements up to zinc, and some molecules \citep[see details in][]{ferland2017}. In total, the model includes up to 625 different species (accounting for atoms and molecules in different ionisation states and isotopes), and the continuity equations are solved individually for each of them. By default, the composition of the gas is assumed to be solar-like, however, the abundances of the specific species can be altered or set to zero (except for hydrogen).
\subsection{Hydrodynamic model}\label{sec::model_hydro}
To model the hydrodynamic outflow, we employ the 1D model presented in \citet{kubyshkina2018grid}, which considers a pure hydrogen atmosphere exposed to XUV stellar irradiation. The code solves the following fluid dynamics equations
\begin{eqnarray}
\frac{\partial\rho}{\partial t} + \frac{\partial(\rho v r^2)}{r^2\partial r} &=& 0\,,\label{eq::mass_cons} \\
\frac{\partial\rho v}{\partial t} + \frac{\partial[r^2(\rho v^2+P)]}{r^2\partial r} &=& - \frac{\partial U}{\partial r} + \frac{2P}{r}\,,\label{eq::moment_cons} \\
\frac{\partial[E_{\rm k}+E_{\rm th}+\rho U]}{\partial t} &+& \frac{\partial vr^2[E_{\rm k}+E_{\rm th}+P+\rho U]}{r^2\partial r} = \nonumber\\
H_{\rm tot} - C_{\rm tot} &+& \frac{\partial}{r^2\partial r}(r^2\chi \frac{\partial T}{\partial r})\,.\label{eq::energy_cons}
\end{eqnarray}
Here, $v$ and $T$ are the bulk velocity and temperature of the escaping atmosphere. The total mass density $\rho$ accounts for different hydrogen species included in the model
\begin{equation}\label{eq::rho_hydro}
\rho = m_{\rm H}(n_{\rm H}+n_{\rm H^+}) + 2m_{\rm H}(n_{\rm H_2}+n_{\rm H_2^+}) + 3m_{\rm H}n_{\rm H_3^+}\,,
\end{equation}
where $n_{\rm i}$ and $m_{\rm i}$ are the number densities and masses of respective species. The atmospheric pressure $P$ is defined as
\begin{equation}
P = (n_{\rm H} + n_{\rm H^+} + n_{\rm H_2} + n_{\rm H_2^+} + n_{\rm H_3^+}+n_{\rm e})kT\,.
\end{equation}
The gravitational potential $U$ accounts for the tidal forces of the host star \citep[Roche lobe effect;][]{Erkaev2007} 
\begin{equation}
\label{eq::gravpot} U = U_0\left[-\frac{1}{\zeta}-\frac{1}{\delta(\lambda-\zeta)}-\frac{1+\delta}{2\delta\lambda^3}\left(\lambda\frac{1}{1+\delta}-\zeta\right)^2\right]\,.
\end{equation}
In Equation\,(\ref{eq::gravpot}), $U_0 = GM_{\rm pl}/R_{\rm pl}$ is the gravitational potential of the planet at the planetary radius \Rpl, which we consider corresponding to the photosphere, $\delta = M_{\rm pl}/M_*$ is the ratio between planetary (\Mpl) and stellar ($M_*$) masses, $\lambda = a/R_{\rm pl}$ is the ratio between the orbital separation ($a$) and the planetary radius, and $\zeta = r/R_{\rm pl}$ is the radial distance from the planetary centre normalised to \Rpl.
In Equation\,(\ref{eq::energy_cons}), $E_{\rm k} = \rho v^2/2$ is the kinetic energy of the outflow, while the thermal energy of the atmosphere is defined as
\begin{equation}\label{eq::thermal_en}
E_{\rm th} = \left[\frac{3}{2}(n_{\rm H}+n_{\rm H^+}+n_{\rm e})+\frac{5}{2}(n_{\rm H_2}+n_{\rm H_2^+})+3\,n_{\rm H_3^+}\right]kT\,.
\end{equation}
In the last term of Equation\,(\ref{eq::energy_cons}), the coefficient
\begin{equation}\label{eq::therm_conduct}
\chi = 4.45\times10^4\left(\frac{T[{\rm K}]}{1000}\right)^{0.7} {\rm erg\,cm^{-1}\,s^{-1}}
\end{equation}
corresponds to the thermal conductivity of the neutral gas \citep{watson1981}, which dominates the thermal conductivity of electrons and ions \citep{kubyshkina2018grid}. Finally, $H_{\rm tot}$ and $C_{\rm tot}$ are the volume heating and cooling rates, respectively. 

In the code presented here, the calculation of $H_{\rm tot}$ and $C_{\rm tot}$, as well as of the whole chemical network, is delegated to Cloudy. However, we present below a brief description of how the hydrodynamic code computes them, because the initial steps of hydrodynamic simulations computed together with Cloudy are based on results of the hydrodynamic code alone (see Section\,\ref{sec::code_base}).

The cooling term in the hydrodynamic code accounts explicitly for {Ly\,$\alpha$ \citep{watson1981} and ${\rm H_3^+}$ cooling \citep[following the approach of ][]{miller2013}}, as described in \citet{kubyshkina2018grid}, and the heating term is exclusively driven by XUV heating. For the input XUV spectrum, the previous version of the code employed the $2\lambda$-approximation that reduces the whole XUV wavelength range to the two specific values of 60\,nm (20\,eV) for the EUV band and 5\,nm (248\,eV) for the X-ray band. This approach allows one to obtain atmospheric escape rates close to those of models employing full spectra {(within a factor of 1.5 for the model planets considered in this study -- see Section\,\ref{sec::gridtests_planets} -- with the largest difference achieved for low XUV values)}, but it distorts the ion fraction and heating profiles \citep{guo_benjaffel2016,odert2020}, hence affecting to some extent the profiles of other atmospheric parameters as well. As Cloudy accounts for the shape of the input spectra, to facilitate the comparison and ease the transition between the two models, we adjusted the hydrodynamic code to employ more detailed spectra and split the incoming XUV flux into $N$ bins according to the specific input SED (see details in Section\,\ref{sec::code_SED}). The specific value of $N$ is set empirically to allow reproducing well a specific spectrum. Namely, we start by employing $N=2$ and increase $N$ by 1 until the differences between the density and temperature profiles predicted by the models employing N and N+1 bins drop below 10\%. 
Therefore, the heating term becomes $H_{\rm tot}$\,=\,$\sum_{m=1}^{N} H_{\rm m}$, where each term represents a specific wavelength (photon energy) interval and takes the form of
\begin{equation}\label{eq::Qxuv}
H_{m} = \eta\,\sigma_{m}\,(n_{\rm H}+n_{\rm H_2})\,\phi_{m}\,.
\end{equation}
The term $\eta$ is the so-called heating efficiency for which we assume a value of 0.15 at all wavelengths, which is a {sufficiently} good approximation for estimating the atmospheric mass-loss rate within the range of planets we are interested in \citep{salz2016}. {We note that, in contrast to the constant parameter $\eta$ used here, the realistic heating efficiency is a height-dependent term \citep[see, e.g.,][]{shematovich2014}. We discuss this in more detail in Section\,\ref{sec::gridtests_HE}.} The term $\sigma_{m}$ is the absorption cross-section for the specific wavelength $\lambda_{\rm m}$ (centre of the m-th bin), while $\phi_{m}$ is the flux absorption function
\begin{equation}
\label{eq::phi_m}
\phi_{m} = \frac{1}{2}\int_{0}^{{\pi}/{2}+\arccos({1}/{r})}\{J_{m}(r,\theta)\sin(\theta) \}\,d\theta\,,
\end{equation}
{where $J_{m}(r,\theta)$ is a function in spherical coordinates describing the spatial variation of the XUV flux of a given wavelength $\lambda_{\rm m}$ due to atmospheric absorption \citep{Erkaev2015}. The absorption cross-section is inversely proportional to the energy of the incoming photon as $\sim E_{\lambda}^{-3}$, therefore it increases about three orders of magnitude from X-ray to EUV part of the spectra. This implies that the stellar X-ray photons penetrate deeper into the planetary atmosphere than the EUV photons, and heat the atmosphere closer to \Rpl, where the density of the atmosphere is higher. Therefore, even though in the given formulation $\phi_{\rm m}$ in the X-ray part (at high photon energies) appears significantly smaller than $\phi_{\rm m}$ in the EUV part,} X-rays can still cause significant atmospheric heating, especially for young planets. The predicted impact of X-ray heating {can increase} further, if one takes into consideration the higher heating efficiency (larger $\eta$ in Equation\,(\ref{eq::Qxuv})) expected at short wavelengths \citep[see e.g.][]{kubyshkina2022_3d}. We discuss this in more detail in Section\,\ref{sec::gridtests}.

Finally, the hydrodynamic model accounts for a range of chemical processes, such as photoionisation and collisional ionisation, recombination of ions, and dissociation of molecular hydrogen. For this purpose, the code solves the set of continuity equations  
\begin{equation}\label{eqn::continuity}
\frac{\partial n_{\rm i}}{\partial t}+\frac{\partial(n_{\rm i}vr^2)}{r^2\partial r} = R_{\rm i} - S_{\rm i}\,, 
\end{equation}
where ``i'' stands for the specific species (molecular and atomic hydrogen, neutral and ionised), while $R_{\rm i}$ and $S_{\rm i}$ represent, respectively, the replenishment and sink of the specific species due to all chemical reactions considered in the model. The complete list of chemical reactions and relative cross-sections considered in the hydrodynamic model is given by \citet{kubyshkina2018grid}. 
\section{Numerical implementation}\label{sec::code} %
In this section, we describe our approach for combining the hydrodynamic code with Cloudy. We start with describing the general scheme of the code (Section\,\ref{sec::code_base}) and then discuss in more detail some of the input parameters and settings applied to the Cloudy runs. In particular, we describe the employed stellar spectra (Section\,\ref{sec::code_SED}), the boundary conditions employed for both Cloudy and the hydrodynamic model (Section\,\ref{sec::code_bord}), the considered atmospheric composition (Section\,\ref{sec::code_muhe}), and the implementation of the adiabatic wind in Cloudy (Section\,\ref{sec::code_wind}).
\subsection{Basic scheme}\label{sec::code_base}
\begin{figure}
	\includegraphics[width=\columnwidth]{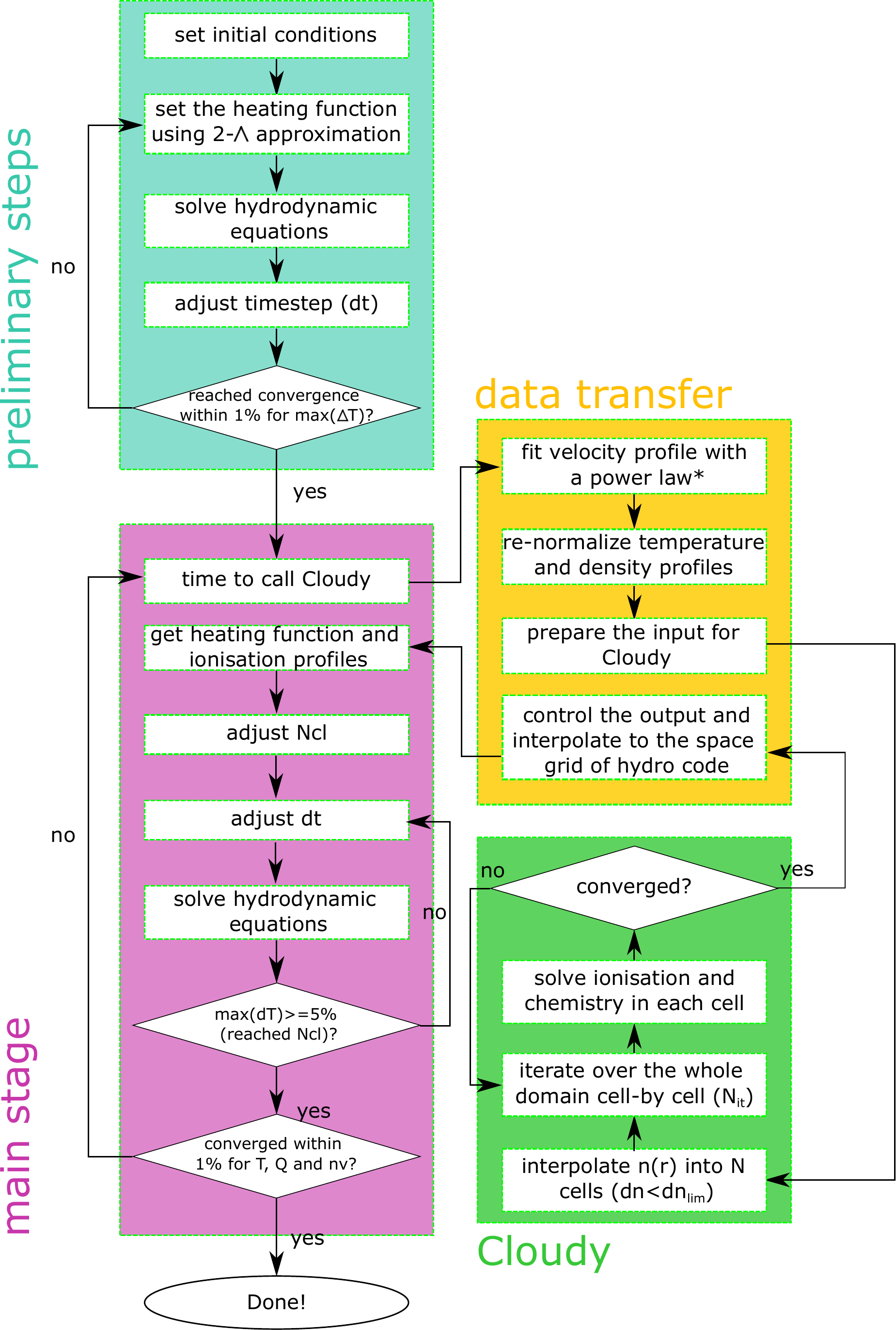}
    \caption{Scheme of the numerical framework. The turquoise area represents the preliminary stage of the simulation carried out by the hydrodynamic code alone. The main stage of the simulation is highlighted in purple, while the yellow and green areas correspond to the data transfer and Cloudy sections. The step marked with an ``*'' in the data transfer section (yellow) is optional (see details in Section\,\ref{sec::code_wind}). See the text for an explanation of the notation.}
    \label{fig::scheme1}
\end{figure}

For the combined framework, we delegate to Cloudy the calculation of $H_{\rm tot}$ and $C_{\rm tot}$, as well as that of the chemical network, while everything else is left to the hydrodynamic code. Figure\,\ref{fig::scheme1} shows the basic scheme of the code. It consists of four main blocks: the preliminary stage (turquoise), the main stage (purple), the block responsible for the data transfer between the hydrodynamic code and Cloudy (yellow), and Cloudy itself (green).

Given the complexity of the physical model and the code, calculations with Cloudy are time-consuming. A single run estimating the heating and cooling terms (hereafter referred to as heating function) and abundances of different atmospheric species, ionised and neutral, across the atmosphere can take from a few minutes (for a simple setup and an atmosphere consisting of pure hydrogen) to a few hours (for an atmosphere including helium and heavier elements, further on referred to as metals{, and wind advection; see the details below}). Therefore, as the hydrodynamic code needs an order of million steps (iterations) to converge, running Cloudy at every step is not feasible.

We tackle this problem in two ways. First, following the approach used previously for the hydrodynamic code \citep[see ][]{Erkaev2016,kubyshkina2018grid}, we expect that the atmospheric structure changes slowly through the iterations of the numerical solution, namely the hydrodynamic timescales are considerably longer than the microphysical timescales (e.g. photoionisation and recombination timescales). Therefore, the heating function (and chemical structure) is redefined once over a certain number of steps $N_{\rm cl}$. When running the hydrodynamic code alone, this number is set to the constant value of 500 steps, which was estimated to be an optimal number for the majority of the cases. Due to the higher complexity of the Cloudy code compared to the heating function described in Section\,\ref{sec::model_hydro}, for the combined framework we apply a different approach. Namely, after the first Cloudy run, we evaluate the maximum relative changes in temperature (i.e. $\Delta T/T$) and in mass flux (i.e. $\Delta(nv)/nv$) across the simulation domain over $N_{\rm cl}$ time steps and then we compare these values to the accuracy constant $C_1$ (typically assigned to 5\%). We then adjust $N_{\rm cl}$ in such a way that 
\begin{equation}\label{eq::Ncloudy}
    \begin{cases}
    \frac{\Delta T}{T}\times\frac{N_{\rm cl,new}}{N_{\rm cl,old}} = C_1\,, & \max(\frac{\Delta T}{T}) > \max(\frac{\Delta(nv)}{nv})\,, \\
    \frac{\Delta(nv)}{nv}\times\frac{N_{\rm cl,new}}{N_{\rm cl,old}} = C_1\,, & \max(\frac{\Delta T}{T}) < \max(\frac{\Delta(nv)}{nv})\,,
    \end{cases}
\end{equation}
and then run Cloudy again. Thus, $N_{\rm cl}$ increases if the relative changes are smaller than $C_1$ to speed up the calculations, and otherwise decreases to keep up to the required accuracy. We then repeat this procedure until the simulation reaches a steady state. As an initial value of $N_{\rm cl}$, we take 10--100 time steps, depending on the type of the simulated planet (the optimal values were defined as part of the code testing). {During the course of testing the code, we verified that varying the initial value of $N_{\rm cl}$ and $C_1$ within reasonable ranges (i.e. 1--1000 steps and 1-20\%, respectively) affects just the number of iterations needed to reach convergence, and not the final solution.}

Second, we include a preliminary stage (the turquoise area in the scheme shown in Figure\,\ref{fig::scheme1}) in our simulations. In this phase, the atmosphere is simulated just by the hydrodynamic code (see Section\,\ref{sec::model_hydro}) until it nearly converges and only then, we start employing Cloudy. In this way, we start the hydrodynamic simulation with Cloudy considering atmospheric parameters that are already realistic and not too far from the final solution. With such an approach, we typically need between {100--800} Cloudy runs until the solution reaches a steady state.

To solve numerically the hydrodynamic equations (Equations\,(\ref{eq::mass_cons}) to (\ref{eq::energy_cons})), both within the preliminary and the main stages, we apply the finite differential McCormack scheme of the second order \citep[predictor-corrector-method; see][ for more detail]{Erkaev2016}.

As we already mentioned in Section\,\ref{sec::model_cloudy}, to solve the photoionisation balance, Cloudy requires as a minimal input the hydrogen density structure as a function of altitude and the SED of the host star. Additionally, one can specify the temperature profile, which crucially depends on the hydrodynamic effects, and information on the wind flow (see details in Section\,\ref{sec::code_wind}). This information, along with the simulation parameters, such as stellar luminosity, orbital distance, and atmospheric composition, has to be passed to Cloudy through input files organised in a specific format. The data transfer block of the code (yellow area in Figure\,\ref{fig::scheme1}) is thus responsible for translating the atmospheric structure data produced by the hydrodynamic code into the format accepted by Cloudy, writing up the Cloudy input files, running Cloudy, control the Cloudy outputs, and, finally, translate the heating function and abundances back to the hydrodynamic code in the correct format.

In essence, the scheme of the code presented here is similar to that of the framework of {\citet[][PLUTO-CLOUDY Interface, TPCI, where PLUTO is the hydrodynamic code by \citealt{Mignone2012}]{salz2015}}, where most of the differences in the technical implementation arise from differences in the underlying hydrodynamic codes. A further difference is in the calculation of the time steps considered above, which is mainly controlled by changes in density and pressure in TPCI, while we use temperature and density. On the physical side, our hydrodynamic code does not include radiative pressure as an external force, but this is expected to have a minor influence on the results in most cases. The major difference between CHAIN and TPCI concerns the basic scheme and namely the inclusion of the preliminary stage in CHAIN allowing to reduce the number of Cloudy runs within one simulation.
\subsection{Stellar spectrum}\label{sec::code_SED}
Cloudy takes the information about the irradiation source in the form of two independent components. One is the shape of the SED (given as a separate input file) and the other one is the intensity of irradiation (set by the total luminosity distributed over the spectrum and the star-planet distance). The spectrum can be defined between low-frequency (low-energy) limit of 10\,MHz up to the high-energy limit of 100\,MeV, which safely covers the infrared to X-ray wavelength range ($\sim0.001-1200$\,eV) most relevant for planetary atmospheres. Furthermore, Cloudy enables the user to include several radiation sources at the same time, so that the total incident radiation represents the sum of their inputs. We use this possibility and split the stellar spectra into three parts corresponding to X-ray (0.5--10\,nm), EUV (10--92\,nm), and UV to infrared (92--3000\,nm\footnote{The upper wavelength boundary given here is indicative and depends on the specific spectrum. For the solar spectrum used in this study, the spectrum is available up to 2400\,nm, while the spectra taken from the MUSCLES survey ({\tt https://archive.stsci.edu/prepds/muscles/}) include fluxes up to 5500\,nm.}) and set the fluxes within these wavelength intervals separately. This approach facilitates the scaling of the XUV flux, which is typically the main driver of the planetary outflow.

In this study, unless specified otherwise, we employ the spectrum of the present-day Sun \citep{Claire2012} and scale the X-ray and EUV fluxes to the desired values (see Sections\,\ref{sec::salz} and \ref{sec::gridtests}). {For use in the hydrodynamic code (within the preliminary stage), we employ only the XUV part of the spectrum, as described in Section\,\ref{sec::model_hydro}; it is sufficient to bin it into 13--15 wavelength intervals to account for the shape of the SED sufficiently well.}
To study the effect of the shape of the stellar spectra on the atmospheric parameters, we additionally consider the spectra provided by the MUSCLES survey \citep[the Measurements of the Ultraviolet Spectral Characteristics of Low-mass Exoplanetary Systems;][]{musclesI,musclesII,musclesIII,musclesIV,musclesV} for three warm Neptunes, namely GJ\,1214\,b \citep{Charbonneau2009,berta2011,narita2013}, GJ\,436\,b \citep{butler2004,knutson2011}, and HD\,97658\,b \citep{Howard2011}. We discuss the results obtained considering the different sources of the stellar SED in Section\,\ref{sec::disscussion_spectra}.
\subsection{Boundary conditions}\label{sec::code_bord}
In the case of 1D models, the simulation domain is set by the conditions imposed at the lower and upper boundaries. For the hydrodynamic model, {we commonly set the lower boundary at the photosphere of the planet \citep[corresponding to pressure levels of $\sim10-1000$\,mbar; see][]{cubillos2017}} and the upper boundary is taken at the Roche radius of the planet ($R_{\rm roche}$), or far enough for the outflow to become supersonic (e.g. for planets at large orbital separations, where the Roche lobe lies at hundreds \Rpl\ from the photosphere). At the upper boundary, we assume a continuous outflow (the first radial derivatives of density, velocity, and temperature are equal to zero), and at the lower boundary, we set pressure and temperature to the photospheric values and assume at the first step that the atmosphere is composed of neutral molecular hydrogen \citep[see details in][]{kubyshkina2018grid}. Throughout the run, after $\rho vr^2$ settles at a constant level in most of the simulation domain, we further control that the mass flow at the lower boundary remains continuous by adjusting the velocity term. In this way, we can avoid a flow discontinuity near the lower boundary originating from the parameters being forced to the fixed values at this point, which appeared in the earlier versions of the hydrodynamic code. We note, however, that this change does not have any considerable effect on the model predictions outside of the small region near the lower boundary (typically $r\lesssim1.05$\,$R_{\rm pl}$ for sub-Neptune-like planets).

We note that setting the lower boundary at the pressure level corresponding to the planetary photosphere can be a matter of debate. We motivate this because we aim at including the whole region in which XUV (or more precisely, X-ray) heating can be effective, and thus account for the whole input from stellar heating. At moderate temperatures ($\sim100-1000$\,K) the atmospheres below $\sim1$\,${\rm\mu}$bar can be strongly affected by condensation (aerosols production), not accounted for by our models. This can affect the opacity of this atmospheric region and restrict or prevent the transport of elements affected by condensation to the upper layers of the atmosphere. We do not expect the former effect to have a large influence on the predictions of our model. Tests of the hydrodynamic model have shown that in most cases the conditions at the lower boundary have a small influence on the results unless the presence of aerosols modifies the temperature at the lower boundary by more than $\sim30$\% or for weakly irradiated low-mass planets. Furthermore, {in most of the cases} the specific pressure at which the lower boundary is set has also a small impact on the results. In the cases in which the heating function below the 1\,${\rm\mu}$bar level is significant (i.e. within an order of magnitude of the peak values), shifting the boundary to lower pressures can lead to {a decrease} of the predicted mass-loss rate and increase of the predicted temperatures, as some portion of the heating is excluded and the atmospheric expansion starts at altitudes with lower density. {For the model planets considered here (see Section\,\ref{sec::gridtests_planets}), we find that placing the lower boundary at pressure levels three orders of magnitude lower than the photospheric values results in mass-loss rate differences of less than a factor of 1.4 for Saturn-like and Earth-like planets, less than a factor of 2.5 for Neptune-like planets, and up to a factor of 5.6 for low-mass planets with voluminous/inflated atmospheres (i.e. super-puffs). In all cases, the maximum differences are achieved for the highest equilibrium temperatures and the highest XUV fluxes, when the atmospheres are most inflated. The changes in maximum temperatures are within $\sim$0.5\% for Saturn-like and Earth-like planets, within $\sim$10\% for Neptune-like planets, and within $\sim$20\% for super-puffs, although in the two latter cases the positions of the temperature maximum can shift upwards by up to a few planetary radii.} Therefore, {except for super-puffs,} heating below the 1\,${\rm\mu}$bar level has a small effect on the predictions of our model. We discuss the impact of condensation on the transport of specific elements in the next sections. %

For what concerns the upper boundary, the continuous outflow conditions employed here force the atmospheric outflow to be hydrodynamic; such treatment becomes inapplicable if the exobase level (i.e. where the atmosphere becomes collisionless) is located in the atmosphere below the sonic point. In this case, atmospheric escape occurs in the Jeans-like regime. In the present study, we verify that the hydrodynamic approach is valid for each of the modelled planets a-posteriori; however, for a broader application, this limitation has to be accounted for within the simulation by switching the upper boundary conditions \citep[e.g. ][]{koskinen2014}.

For the combined framework, we have to slightly alter {the approach used by the hydrodynamic code}. The typical densities at the photosphere predicted by lower atmosphere models are of the order of $10^{18}$\,cm$^{-3}$, while Cloudy has a maximum density limit of $10^{15}$\,cm$^{-3}$ above which calculations may become unreliable. Therefore, {for modelling the entire atmosphere, the code is designed to employ} the solution of the hydrodynamic code where the density is higher than $10^{15}$\,cm$^{-3}$ and consider what Cloudy gives at lower densities, controlling that the solution is continuous. Otherwise, when focusing on the upper atmosphere, we set the lower boundary directly at $10^{15}$\,cm$^{-3}$ {or slightly above this level}. %
Furthermore, the Roche lobes of close-in sub-Neptunes can lie close to the planets, namely just a few \Rpl\ away from the photosphere, and placing the upper boundary so close to the planet can lead to numerical problems in Cloudy. Therefore, for the combined framework we ensure that the upper boundary stands always at least 15\,\Rpl\ away from the planet.

\subsection{Atmospheric composition}\label{sec::code_muhe}
As discussed in Section\,\ref{sec::model_cloudy}, Cloudy can take into account elements from hydrogen to zinc and by definition assumes solar composition. However, the abundance of specific elements can be adjusted by the user and each specific element, except for hydrogen, can be excluded from the simulation. This provides significant flexibility and is most relevant for hot planets with massive atmospheres (see Section\,\ref{sec::gridtests_muhe}).

In this study, we consider the following three types of atmospheres: 
\begin{enumerate}
    \item pure hydrogen atmosphere including atomic and molecular hydrogen species, 
    \item hydrogen-helium atmosphere with helium fraction of 10\%, and
    \item atmosphere including all elements up to zinc with solar abundances.
\end{enumerate}
In the first case, the mean molecular weight of the atmosphere is set by hydrogen, as in the hydrodynamic code described in Section\,\ref{sec::model_hydro}. Therefore, the differences in predicted atmospheric parameters between this case and the hydrodynamic model are purely given by a more accurate prescription of the hydrogen chemistry and, hence, a more realistic heating function. The better resolution of the stellar spectra described above (in comparison to the $2\lambda$-approximation used by the hydrodynamic model) affects mainly the temperature and ionisation profiles across the atmospheres but has, in general, a minor effect on the outflow parameters, as we show in Section\,\ref{sec::gridtests_muhe}. 

Differently from case (i) described above, the inclusion of a substantial fraction of helium in cases (ii) and (iii) affects considerably the mean molecular weight of the atmosphere, which can, in turn, affect the basic parameters of the hydrodynamic outflow. Therefore, when Cloudy is called in the framework, we adjust the mean molecular weight also in the hydrodynamic part of the code according to that of the hydrogen-helium mixture. For Neptune-like planets, this adjustment has a small but non-negligible impact on the results, which often appears more significant than the additional heating/cooling due to the chemical reactions involving helium. We will discuss this in more detail in Section\,\ref{sec::gridtests}. Finally, in case (iii) we assume that the fraction of metals in the atmosphere relative to hydrogen and helium is too small ($\sim0.1$\%) to affect considerably the mean molecular weight of the atmosphere and, thus, to affect the hydrodynamic outflow.
\subsection{Adiabatic wind}\label{sec::code_wind}
Although the hydrodynamic solver is not included in Cloudy, it still enables users to account to some extent for gas motion, and thus for a planetary wind. More specifically, the recent versions of Cloudy \citep{ferland2013,ferland2017} include the possibility to prescribe a gas flow set up by an external model, which is then accounted for in the calculations of the advection of the atmospheric species (i.e. of the source and sink terms at the boundaries between the cells of the simulation domain). To this end, the flow ($\Phi = nv$) across the atmosphere has to be parameterised by the power law describing where in the atmosphere the wind is initiated ($r_0$) and which velocity the flow reaches at the upper boundary of the simulation domain ($v_1$). Therefore, to account for the planetary wind effects we approximate the flow calculated by the hydrodynamic code (within the data transfer section of our code) each time before the Cloudy run as
\begin{equation}
    nv = \Phi_1 \left( \frac{(r-r_0)}{(r_1 -r_0)}\right)^l\,.
\end{equation}
Here, $\Phi_1 = n_1v_1$ is the flow at the upper boundary (given that the density profile is already set, the code passes just the $v_1$ value to Cloudy) and $l$ is the approximated power law index. In general, the approximation error does not exceed a few percent, and the maximum difference is reached at the region where the bulk flow is initiated and the outflow velocity is close to zero. Given that only relatively large velocities (order of a few km\,s$^{-1}$) can have a significant influence on the advection, such accuracy is sufficient.

In both cases, with or without the wind prescription, Cloudy applies the iterative steady-state solver and goes through the simulation domain cell by cell until the solution converges. However, the number of iterations needed to reach convergence depends on whether the wind is included or not when running Cloudy. For simple hydrogen and hydrogen-helium atmospheres, the simulation without wind prescription requires just a few ($\sim$3--5) iterations to converge, while the one with wind prescription needs a few tens of iterations for the same model. Therefore, to reduce the total simulation time we do not include the wind from the beginning. Instead, we let the model nearly converge and only then ``switch on'' the wind option in the Cloudy simulations. It is a reasonable approach, as for most of the planets the wind advection has just a minor impact on the results {(see details in Section\,\ref{sec::gridtests_grid} and Figure\,\ref{fig::comp_hydro_H})} and the ``no-wind'' solution is typically close to the final one. {Technically, this is equivalent to starting the simulation from different initial profiles, which has an effect on the number of iterations needed to reach convergence, but not on the final results. We verified this for the original hydrodynamic model and the combined CHAIN model.}

Finally, we note that the wind advection remains an experimental feature of the Cloudy code and it is still under development (see code manuals at {\tt gitlab.nublado.org/cloudy/cloudy/-/wikis/home}). This means that this implementation can suffer from numerical problems (particularly in the pressure solver when passing through the sonic point), some of which we have encountered through the testing phase of this study. Therefore, we consider the ``no-wind'' version of the model as the default one and treat the inclusion of the wind in the Cloudy simulations as an optional feature. 

\section{Validation of the model}\label{sec::salz}
To validate our framework, we compare the results of simulations of nine specific planets with previously published work by \citet{salz2016a}, which used a very similar approach to the one employed here \citep{salz2015}, though considering an earlier version of Cloudy \citep{ferland2013}. Specifically, we compare the results for five sub-Neptune-like planets (GJ\,3470\,b, HAT-P-11\,b, GJ\,1214\,b, GJ\,436\,b, and HD\,97658\,b) and four hot Jupiters (HD\,149026\,b, HD\,189733\,b, HD\,209458\,b, and WASP-80\,b). {We stress that the simulations performed here do not aim at describing the realistic physical conditions of these planets, where the assumptions used by our models might be not valid \citep[e.g. according to recent observations, the atmosphere of GJ\,1214\,b is rich with metals and affected by cloud formation][]{kempton2023}, but we only use them to compare to older results to test the code. Thus, any discussion provided here considers only the behaviour of the model.} For this test, we employ the same planetary and stellar parameters as in \citet{salz2016a} and set the lower boundary near the high-density applicability limit of Cloudy (i.e. $10^{15}$\,cm$^{-3}$). The latter corresponds to atmospheric pressures of the order of $\sim10^{-5}-10^{-4}$\,bar, which is much smaller than the photospheric pressures ($\sim10^{-2}-10^{-1}$\,bar) that we typically employ when running the hydrodynamic code alone. For simplicity, we adopt this approach throughout this study. %

The atmospheric composition adopted in \citet{salz2016a} is a hydrogen-helium mixture with solar relative abundances and therefore corresponds to the case (ii) described in Section\,\ref{sec::code_muhe}. The models of \citet{salz2016a} also include wind advection, which is discussed in Section\,\ref{sec::code_wind}. For the comparison, we consider the results we obtained both with and without accounting for wind advection in the Cloudy simulations.%

We follow the approach to construct the stellar spectra similar to that used by \citet{salz2016a}. We scale the X-ray, EUV, and bolometric luminosities to the same values \citep[adopted from ][]{dere1997chianti,dere2009chianti,pizzolato2003,linsky2014} and use the same spectral resolution as in \citet{salz2016a}. The main difference with \citet{salz2016a} is that for the shape of the SED, we use the solar spectrum given by \citet{Claire2012}, while \citet{salz2016a} uses that of \citet{woods_rot2002} for the XUV band and adopts black body radiation at longer wavelengths. Additional minor differences with the work of \citet{salz2016a} are in the setup of the boundary conditions as a result of differences in the underlying hydrodynamic models.
\begin{table*}[t]
	\centering
	\caption{Parameters of the planets considered for the comparison with the results of \citet{salz2016a}. The last three columns correspond to the atmospheric mass-loss rate obtained by \citet[][ $\dot{M}_{\rm Salz2016}$]{salz2016a} and obtained with our model including ($\dot{M}_{\rm wind}$) and excluding ($\dot{M}_{\rm def}$) wind advection in the Cloudy computation.}
	\label{tab::salzplanets}
	\begin{tabular}{lcccccccccr} 
		\hline
		planet & \Mpl & \Rpl & \Teq & $a$ & $M_*$ & $F_{\rm X}$ & $F_{\rm EUV}$ & $\dot{M}_{\rm Salz2016}$ & $\dot{M}_{\rm wind}$ & $\dot{M}_{\rm def}$\\
		      & [\Mer] & [\Rer] & [K] & [AU] & [\Msun] & [${\rm erg/s/cm^2}$] & [${\rm erg/s/cm^2}$] & [${\rm g/s}$] & [${\rm g/s}$] & [${\rm g/s}$] \\
		\hline
  GJ\,3470\,b & 14.0 & 4.14 & 650 & 0.036 & 0.539 & 1170.4 & 6431.6 & $1.86\times10^{11}$ & $1.51\times10^{11}$ & $2.10\times10^{11}$\\
  
  HAT-P-11\,b & 26.4 & 4.70 & 850 & 0.053 & 0.8 & 449.1 & 2706.3 & $8.08\times10^{10}$ & $8.01\times10^{10}$ & $8.69\times10^{10}$\\
  
  GJ\,1214\,b & 6.36 & 2.69 & 550 & 0.014 & 0.18 & 147.5 & 739.0 & $1.98\times10^{10}$ & $2.05\times10^{10}$ & $2.54\times10^{10}$\\
  
  GJ\,436\,b & 23.2 & 4.25 & 650 & 0.029 & 0.45 & 38.6 & 583.6 & $1.88\times10^{10}$ & $1.53\times10^{10}$ & $1.56\times10^{10}$\\
  
  HD\,97658\,b & 7.95 & 2.35 & 750 & 0.08 & 0.85 & 92.2 & 860.5 & $1.23\times10^{10}$ & $1.12\times10^{10}$ & $1.50\times10^{10}$\\
  
  HD\,149026\,b & 114.4 & 7.28 & 1440 & 0.043 & 1.3 & 7655.8 & 12134 & $1.10\times10^{11}$ & $1.11\times10^{11}$ & $1.50\times10^{11}$\\
  
  HD\,189733\,b & 349.8 & 12.31 & 1200 & 0.031 & 0.81 & 5600.2 & 15073 & $1.59\times10^{10}$ & $1.39\times10^{10}$ & $2.05\times10^{10}$\\
  
  HD\,209458\,b & 219.4 & 15.67 & 1320 & 0.047 & 1.12 & 40.4 & 1113.6 & $8.16\times10^{10}$ & $1.67\times10^{11}$ & $2.62\times10^{11}$\\
  
  WASP-80\,b & 174.88 & 10.64 & 800 & 0.034 & 0.57 & 2177.6 & 8870.9 & $1.46\times10^{11}$ & $1.92\times10^{11}$ & $2.18\times10^{11}$\\  
\hline
	\end{tabular}
\footnotesize{\\ \emph{Note:} According to \citet{salz2016a}, the planetary and stellar parameters were adopted from the following studies. GJ\,3470\,b: \citet{bonfils2012,biddle2014} ; HAT-P-11\,b: \citet{bakos2010}; GJ\,1214\,b: \citet{Charbonneau2009,berta2011}; GJ\,436\,b: \citet{butler2004,knutson2011}; HD\,97658\,b: \citet{Howard2011}; HD\,149026\,b: \citet{sato2005,knutson2009}; HD\,189733\,b: \citet{Bouchy2005,henry_winn2008,Southworth2010,knutson2009}; HD\,209458\,b: \citet{Charbonneau2000,henry2000,torres2008,silva-valio2008,Crossfield2012}; and WASP-80\,b: \citet{triaud2013}. The values of atmospheric mass-loss rates given here are a factor of four larger compared to those given in Table\,3 of \citet{salz2016a}, as we do not apply surface averaging.}
\end{table*}

The most relevant planetary parameters and the atmospheric mass-loss rates given by \citet{salz2016a} and obtained from our model are listed in Table\,\ref{tab::salzplanets}. 
To avoid possible biases, the mass-loss rates $M_{\rm Salz2016}$ listed in Table\,\ref{tab::salzplanets} were computed from the density and velocity profiles published by \citet{salz2016a}, instead of adopting them from their Table\,3. 
We find an excellent agreement between the mass-loss rates obtained by \citet{salz2016a} and from our model accounting for wind advection ($\dot{M}_{\rm Salz2016}$ and $\dot{M}_{\rm wind}$, respectively). The mass-loss rates obtained without wind advection ($\dot{M}_{\rm def}$) are, instead, $\sim$5--50\% higher, but such a difference is negligible in terms of planetary atmospheric evolution.
\begin{figure}
    \centering
    \includegraphics[width=\columnwidth]{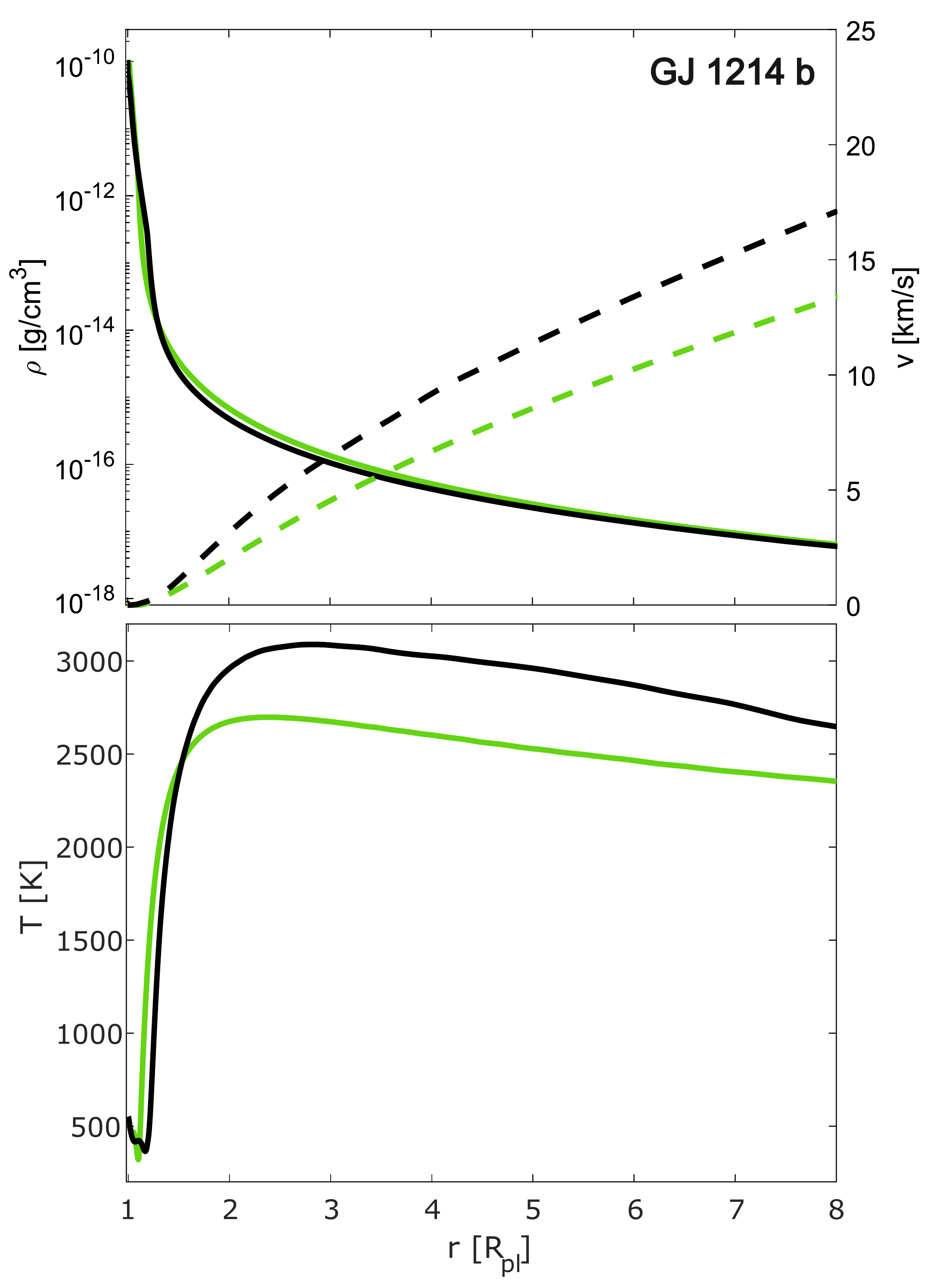}
    \caption{Model results obtained for GJ\,1214\,b. Top: density (left axis, solid lines) and velocity (right axis, dashed lines) profiles. Bottom: temperature profiles. In both panels, green lines are for the results obtained by \citet{salz2016a}, while the black lines are the results obtained from our model with wind advection in the Cloudy computations.}
    \label{fig::salz_compare}
\end{figure}

Figure\,\ref{fig::salz_compare} shows a comparison between the atmospheric profiles obtained by our model computed with %
wind advection and those of \citet{salz2016a}. The density profiles are nearly identical and the small differences below $r\sim2-2.5$\,\Rpl\ are due to differences in the hydrodynamic models. The bulk outflow velocity at the Roche radius ($\sim4$\,\Rpl) and beyond we obtained are $\sim10\%$ larger than that predicted by \citet{salz2016a}. Similarly, we obtained about 10\% higher maximum atmospheric temperature compared to \citet{salz2016a}. These differences are connected and are both caused by differences in the heating in the lower atmosphere, which are likely due to differences in the employed stellar SEDs (see the discussion in Section\,\ref{sec::disscussion_spectra}).

\section{Application to Neptune-like planets}\label{sec::gridtests}
\subsection{Model planets}\label{sec::gridtests_planets}
To test how the inclusion of Cloudy in the computation of the energy balance affects the predicted atmospheric parameters of (sub-)Neptune-like planets, we consider here a set of planets with significantly different masses, radii, and orbital semi-major axes, and under different levels of stellar irradiation. We consider the following four cases:
\begin{enumerate}
    \item[(1)] massive planet hosting a large atmosphere, with a mass of 45\,\Mer\ and a radius of 4\,\Rer, further on referred to as ``sub-Saturn'';
    \item[(2)] Earth-like planet ($M_{\rm pl} = 1M_{\oplus}$, $R_{\rm pl} = 1R_{\oplus}$) with a thin hydrogen-dominated atmosphere, further on referred to as ``Earth-like planet'';
    \item[(3)] low-mass planet of 2.1\,\Mer\ with a large hydrogen-dominated atmosphere ($R_{\rm pl} = 2\,R_{\oplus}$), further on referred to as ``super-puff'';
    \item[(4)] a typical sub-Neptune-like planet with a mass of 5\,\Mer\ and a radius of 2\,\Rer, further on referred to as ``sub-Neptune''.
\end{enumerate}

{These test planets were chosen to represent the range of cases relevant for evolution studies of low- and intermediate-mass planets, which are those most vulnerable to atmospheric loss. Planet (1) represents the low-mass edge of the sub-giant regime, where planets are generally stable against photoevaporation, but can still lose a significant fraction of their mass if located close enough to their host stars \citep[e.g.][]{hallat2022subsaturns}. Planet (2) lies close to the terrestrial-like regime and is not expected to keep its hydrogen-dominated atmosphere for long \citep[e.g.][]{kubyshkina2022MR}. Planet (3) has parameters that are expected to be typical of very young low-mass planets \citep[e.g.][]{lopez_fortney2013}; mass loss for such planets can be, to a large extent, controlled by their high internal thermal energy and low gravity, and in case only slightly enhanced by XUV-heating \citep[e.g.,][]{kubyshkina2018grid,gupta_schlichting2019,owen2023arXiv230800020O}. Finally, planet (4) represents a typical sub-Neptune-like planet close to the boundary between the sub-Neptune and super-Earth populations \citep[e.g.][]{fulton2017}. The stability of the atmospheres of such planets strongly depends on their orbital separation and host star properties \citep[e.g.][]{kubyshkina2022MR}.}

{The atmospheric composition of these model planets is assumed to be hydrogen-dominated (for each, we consider the range of compositions described in Section\,\ref{sec::code_muhe}). The estimates of exact masses and extension of their lower atmospheres depend strongly on the assumptions of the internal structure models (e.g. the temperature and composition of the planetary solid core and the presence of water layers). In general, one can expect that the atmospheric mass fraction of the sub-Saturn model planet (1) is about 10-20\%, and its extension is about 1.5\,$R_{\oplus}$ (assuming a ``rocky'' -- silicate plus iron -- core of $\sim$2.5\,$R_{\oplus}$ and core temperatures typical for Gyr-old planets). Instead, for the Earth-like planet (2), the hydrogen-dominated atmosphere can only be $\leq\sim$0.1\% of the total planetary mass, and its extension is negligible. For both the super-puff (3) and the sub-Neptune (4), the atmospheric mass fraction is of a few percent: in this mass range, the atmospheric mass and the temperature of the core are strongly degenerate for a given radius (e.g. there can be less atmosphere with a hotter core or more atmosphere with a cooler core, with a difference in atmospheric mass of at least a few times). Assuming core radii of $\sim$1.5 and $\sim$1.2\,$R_{\oplus}$, respectively, the atmosphere extends to $\sim$0.5 and $\sim$0.8\,$R_{\oplus}$. 
The above estimates were made based on the rocky core mass-radius relations by \citet{rogers_seager2010} and the lower atmosphere structure models derived by MESA \citep[Modules for Experiments in Stellar Astrophysics;][]{paxton2018}. However, these estimates can vary significantly between different structure models. Nevertheless, the predictions of our upper atmosphere models do not depend on the properties of the lower atmospheres other than the photospheric parameters, which have a small effect on our results (see Section\,\ref{sec::code_bord}).}

We consider {planets (1)--(4)} at different orbital separations around a solar-mass star corresponding to equilibrium temperatures of 2000, 1500, and 1000\,K in the case of the sub-Saturn, and of 1500, 1000, and 700\,K for the other three cases. At the given orbits, the planets (1)--(4) are expected to experience an intense XUV-driven hydrodynamic outflow. At higher equilibrium temperatures compared to those considered here, the atmospheric escape from the planets (2)--(4) can be {fully driven by the internal thermal energy with XUV heating becoming irrelevant} \citep[see e.g.][]{gupta_schlichting2019,kubyshkina2021mesa}. At much lower temperatures, instead, the escape becomes dominated by kinetic/non-thermal processes, especially for low levels of stellar irradiation \citep[see e.g.][]{gronoff2020}.
\begin{figure}
    \centering
    \includegraphics[width=\columnwidth]{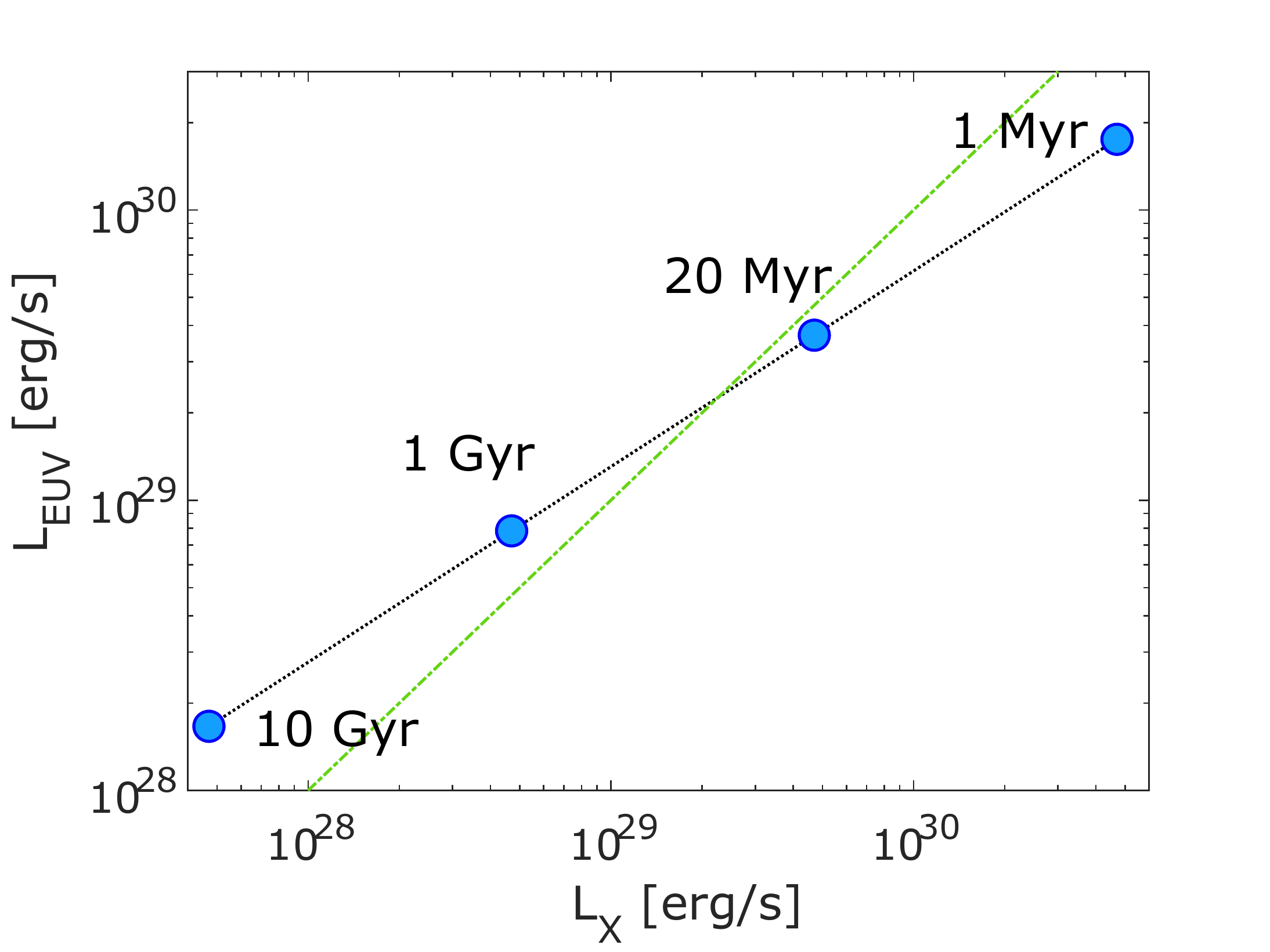}
    \caption{Stellar X-ray and EUV luminosities adopted by the models described in Section\,\ref{sec::gridtests} as given by the Mors code \citep[blue circles on the black line;][]{johnstone2021mors}. The green line in the background denotes for reference the $L_{\rm X} = L_{\rm EUV}$ level.}
    \label{fig::Lxuv_mors}
\end{figure}

Atop of the above, we consider four different activity levels of the host star. To model the intensity of the stellar XUV emission, we employ the Mors code \citep{johnstone2021mors,spada2013} and adopt the stellar X-ray and EUV luminosities predicted for the moderately rotating solar-mass star at ages of about 1\,Myr, 20\,Myr, 1\,Gyr, and 10\,Gyr. The first two ages roughly cover the possible range of protoplanetary disk dispersal times \citep[e.g.][]{mamajek2009}, after which the evaporation of planetary atmospheres begins. The age of 1\,Gyr corresponds approximately to the end of the initial phase of extreme atmospheric escape in planetary evolution \citep[e.g.][]{kubyshkina2020mesa}, while the last point of 10\,Gyr is near the end of the main sequence phase of the host star. Between two consecutive points in the four considered ages, the X-ray luminosity changes by a factor of ten, and in total it varies from $4.7\times10^{30}$\,erg\,s$^{-1}$ at 1\,Myr and $4.7\times10^{27}$\,erg\,s$^{-1}$ at 10\,Gyr. Following the empirical approximations used in the Mors model, the relation between stellar X-ray and EUV luminosity changes with time as illustrated in Figure\,\ref{fig::Lxuv_mors}. At the age of 1\,Myr, the stellar high-energy radiation is X-ray dominated ($F_{\rm X} \simeq 1.7F_{\rm EUV}$). At 20\,Myr, the two luminosities become nearly equal, while at the age of 10\,Gyr this relation evolves to $F_{\rm X} \simeq 0.28F_{\rm EUV}$. Therefore, between two consecutive points in the four considered ages, the EUV luminosity changes roughly 4.7 times. Scaled to the orbital separations corresponding to planetary equilibrium temperatures in the 700--2000\,K range, the total XUV flux lies in the ${220.6-(4.5\times10^6)}$\,erg\,s$^{-1}$\,cm$^{-2}$ range. This is sufficient to ensure that the model planets undergo hydrodynamic escape and test how the photoionisation functions perform under different irradiation levels.

Table\,\ref{tab::gridplanets} in Appendix gives the full list of modelled planets. For each of them, we simulate the atmosphere employing the following configurations of our model: pure hydrodynamic code described in Section\,\ref{sec::model_hydro} (referred to in the following as ``model\,\#0''); combined hydrodynamic and Cloudy model assuming pure hydrogen atmosphere (see Section\,\ref{sec::code_muhe}) and not accounting for the wind advection (``model\,\#1''); combined hydrodynamic and Cloudy model assuming pure hydrogen atmosphere and including wind advection (``model\,\#2''); combined hydrodynamic and Cloudy model for a hydrogen-helium atmosphere and without wind advection (``model\,\#3''); combined hydrodynamic and Cloudy model for a hydrogen-helium atmosphere and including wind advection (``model\,\#4''); and combined hydrodynamic and Cloudy model considering metals and excluding wind advection (``model\,\#5''). %

To ease the understanding of our results, we start by analysing the output of different model configurations for a specific planet. In particular, we compare the results obtained with the different model assumptions listed above for the hot sub-Neptune under moderate XUV irradiation corresponding to the 1\,Gyr-old host star. Next, in Section\,\ref{sec::gridtests_grid} we compare the estimates of the basic atmospheric parameters obtained considering a pure hydrogen atmosphere as given by the hydrodynamic model (model\,\#0) and models\,\#1 and \#2 for the entire set of synthetic planets described above. Finally, we discuss the effects of different atmospheric compositions in Section\,\ref{sec::gridtests_muhe}. 

{For all simulations considered in the following sections, we performed our standard set of diagnostics to verify if the model results are physically adequate. In particular, we control that the exobase level lies above the sonic point or outside of the simulation domain (hence, the hydrodynamic approach is valid), the mass flow $\rho vr^2$ is constant throughout the simulation domain, and the terms of the energy equation are in balance.}
\subsection{Atmospheric outflow from the hot sub-Neptune}\label{sec::gridtests_HE}
\begin{figure}
    \centering
    \includegraphics[width=0.97\columnwidth]{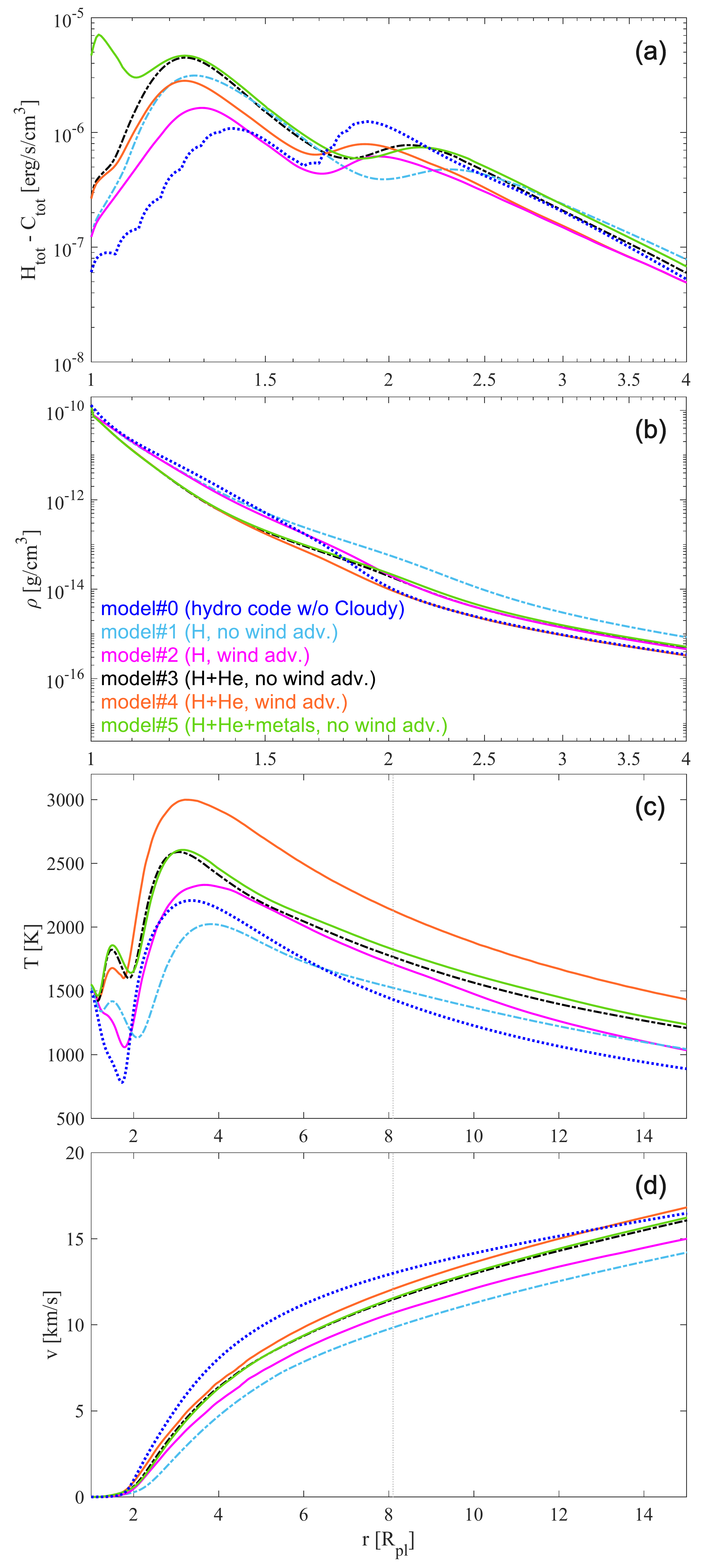}
    \caption{Atmospheric profiles predicted for the hot (1500\,K) 5\,\Mer\ sub-Neptune under moderate XUV irradiation (model planet 4.2B in Table\,\ref{tab::gridplanets}). Different line styles correspond to different model configurations, as indicated in the legend located in panel (b). From top to bottom, the panels show profiles for volume heating rate (a), {mass density} (b), temperature (c), and bulk outflow velocity (d) as a function of radial distance. In the case of the basic hydrodynamic code (model\,\#0), the volume heating rate corresponds to the sum of the input from the X-ray and EUV heating functions given in Section\,\ref{sec::model_hydro} ($H_{\rm X} + H_{\rm EUV}$). Note, that the maximum distance $r = 4$\,\Rpl\ shown in panel (a) is smaller than that considered in the other panels to facilitate the inspection of the details of the heating functions. The vertical grey line in panels (b), (c), and (d) denotes the position of the Roche lobe. The 10$^{15}$\,cm$^{-3}$ high-density limit of Cloudy lies at the lower boundary.}
    \label{fig::6_2b_profiles}
\end{figure}

We start by considering the outputs of models\,\#0 to \#5 for the sub-Neptune planet orbiting its (solar-mass) host star at a distance of 0.0404\,AU (corresponding to an equilibrium temperature of 1500\,K) and exposed to the moderate XUV irradiation level of $\sim2.7\times 10^4$\,erg\,s$^{-1}$\,cm$^{-2}$ (corresponding to a stellar age of 1\,Gyr; model planet 4.2B in Table\,\ref{tab::gridplanets}). In Figure\,\ref{fig::6_2b_profiles}, we compare the atmospheric profiles for volume heating rate, mass density, temperature, and bulk outflow velocity predicted by the six models, while in Table\,\ref{tab::6_2B_cmodels} we compare some of the basic outflow parameters, namely mass-loss rate ($\dot{M}$), maximum atmospheric temperature ($T_{\rm max}$), outflow velocity and density at the Roche radius ($V_{\rm roche}$ and $\rho_{\rm roche}$), sonic radius ($r_{\rm sound}$), and total ion fraction {defined as}%
\begin{equation}\label{eq::fionint}
    {\int f_{\rm ion} = \int_{Rpl}^{Rroche}(n_{\rm ion}/n_{\rm neu})dr\,.}
\end{equation}
\begin{table*}
        \centering
	\caption{Basic parameters of the atmospheric outflow predicted by different model configurations for the hot sub-Neptune planet under moderate XUV irradiation (model planet 4.2B in Table\,\ref{tab::gridplanets}).}
	\label{tab::6_2B_cmodels}
        \begin{tabular}{l|c|c|c|c|c|c|r}
        \hline
            model & comment & $\dot{M}$ & $T_{\rm max}$ & $V_{\rm roche}$ & $\rho_{\rm roche}$ & $r_{\rm sound}$ & $\int(f_{\rm ion})$ \\ 
            \#  & & [10$^{10}$g/s] & [K] & [km/s] & [10$^7$g/cm$^3$] & [\Rpl] & [10$^{-4}$] \\
             \hline
             \#0 & hydro code without Cloudy & {9.03} & {2209} & {13.0} & {3.092} & {2.74} & {18.38} \\
             \#1 & H-atmosphere, no wind advection & 13.05 & 2023 &  9.8 & 5.900 & 3.71 &  4.65 \\
             \#2 & H-atmosphere, wind advection &  8.39 & 2343 & 10.7 & 3.475 & 3.46 &  2.70 \\
             \#3 & H$ + $He, no wind advection &  9.99 & 2589 & 11.5 & 3.871 & 3.26 &  5.53 \\
             \#4 & H$ + $He, wind advection &  7.23 & 3003 & 12.1 & 2.666 & 3.27 &  3.50 \\
             \#5 & H$ + $He$ + $metals, no wind advection & 10.73 & 2606 & 11.6 & 4.140 & 3.31 &  5.48 \\
             \hline
        \end{tabular}
\end{table*}

Figure\,\ref{fig::6_2b_profiles} demonstrates that the shape of the atmospheric profiles and the position of the maxima/minima is similar across all models. {However, models\,\#1 to \#5 predict higher densities, lower velocities, closer-in maxima of the heating function, and, in general, higher temperatures of the outflow compared to model\,\#0. The wind advection in models\,\#2 and \#4 is responsible for the redistribution of the (neutral and ionised) species upwards and adiabatic cooling. This leads to the reduction of the heating in the lowermost layers of the atmosphere (which can be seen well in volume heating and temperature profiles in panels (a) and (c) of Figure\,\ref{fig::6_2b_profiles}, compare the magenta and light-blue lines or orange and black lines) and the consequent reduction in the outflow density and increase in the outflow velocity.}
\subsubsection{Pure hydrogen atmospheres}
\begin{figure}%
    \centering
    \includegraphics[width=\columnwidth]{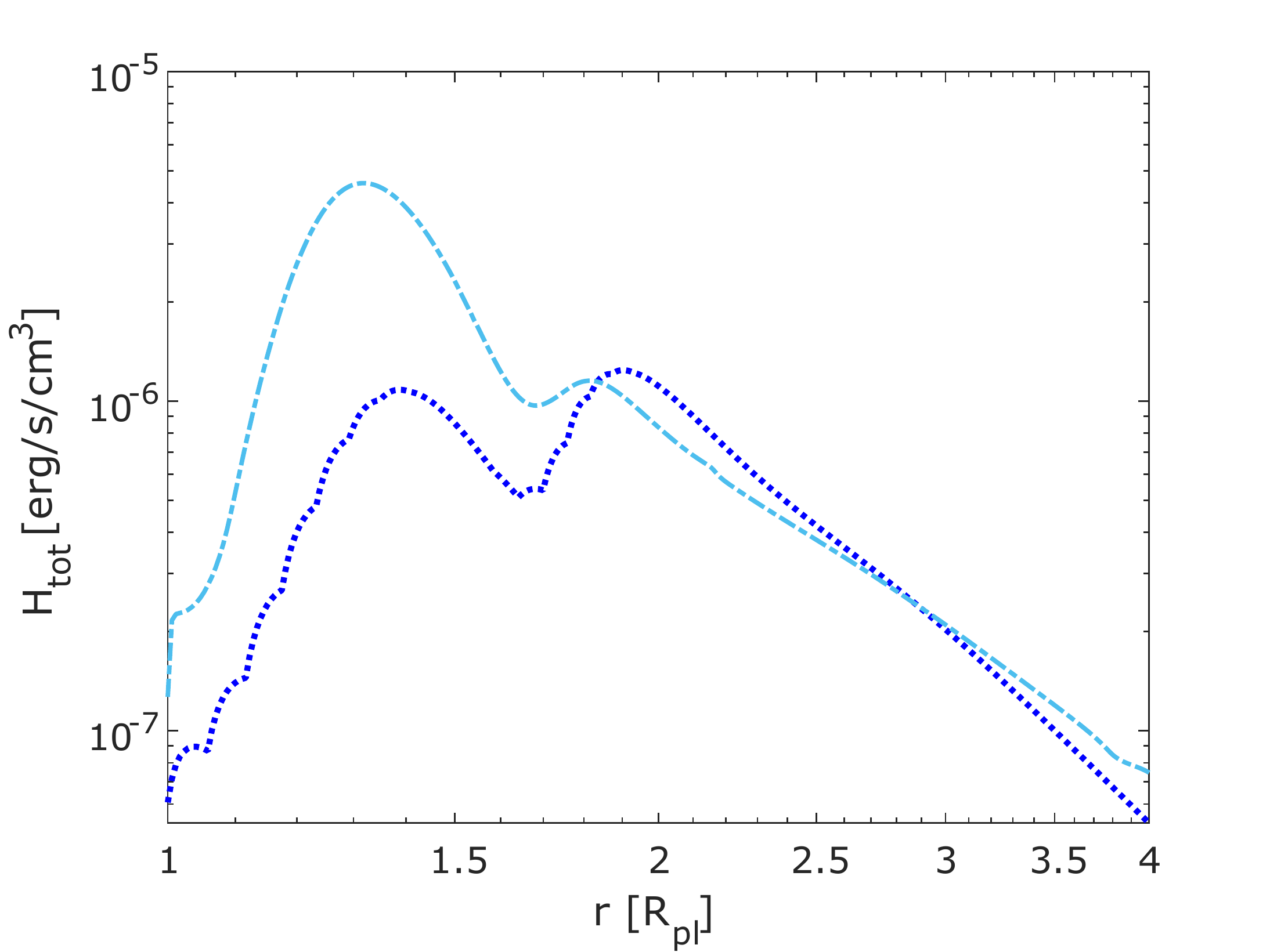}
    \caption{Volume heating rate predicted by the basic hydrodynamic model\,\#0 (dark-blue dotted line) compared to the volume heating rate calculated considering Cloudy for the same density and temperature profiles (light-blue dashed-dotted line), not accounting for wind advection. These profiles result following the first instance in which Cloudy is run in the code.}
    \label{fig::6.2b_1st}
\end{figure}

The direct comparison of the results of the hydrodynamic model described in Section\,\ref{sec::model_hydro} (model\,\#0) with those obtained employing the more precise Cloudy calculations can be best done by considering the pure hydrogen atmosphere (model\,\#1 and model\,\#2). {Model\,\#1 predicts mass-loss rates of $\sim1.45$ times higher and the density of the outflow at the Roche radius is almost a factor of two larger than those given by model\,\#0. Instead, the bulk outflow velocity and the maximum temperature obtained from model\,\#1 are $\sim1.3$ and $\sim1.1$ times smaller than those predicted by model\,\#0, respectively (see Table\,\ref{tab::6_2B_cmodels}). These differences decrease with the inclusion of wind advection (model\,\#2), thus employing physics closer to that of model\,\#0, but the inclusion of Cloudy leads to a slower and denser outflow, a slightly lower atmospheric mass-loss rate, and a slightly higher maximum temperature of the atmosphere (the latter two are not the case for all planets, as shown later). }
\begin{figure}
    \centering
    \includegraphics[width=\columnwidth]{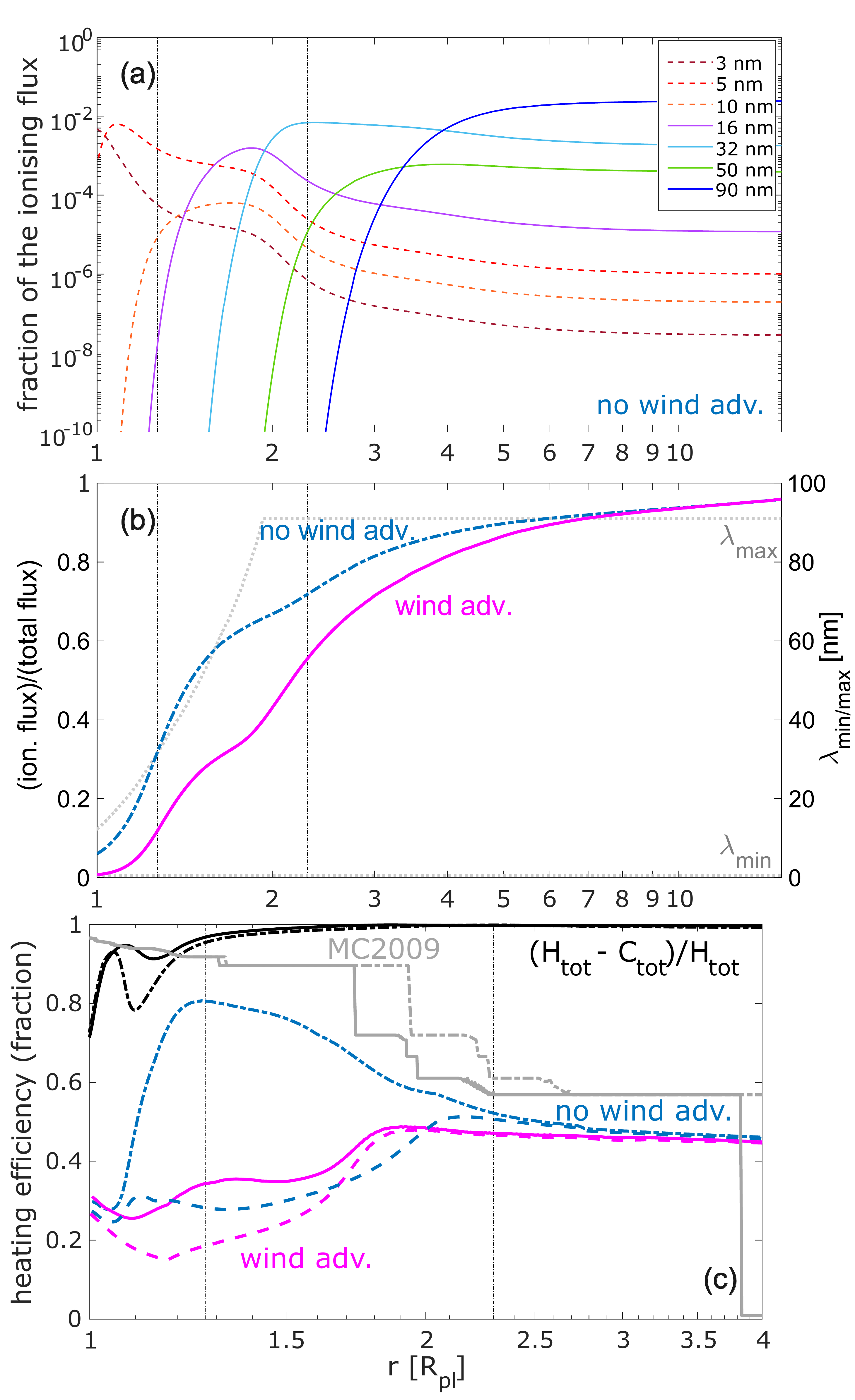}
    \caption{Details of the heating profiles obtained from models\,\#1 and \#2 for the hot sub-Neptune 4.2B. Panel (a): fraction of total photoionisation rate corresponding to the specific irradiation wavelengths as a function of radial distance. The dashed lines are for wavelengths in the X-ray band, while the solid lines are for wavelengths in the EUV band (see the legend). Panel (b): fraction of absorbed stellar flux (blue and magenta lines corresponding to model\,\#1 and model\,\#2, respectively; left y-axis) and total span of absorbed wavelengths at each given radial distance (grey dotted lines; right y-axis). Panel (c): heating efficiency following the last Cloudy iteration of model\,\#1 (blue dashed-dotted line), and the last Cloudy iteration of model\,\#2 (magenta solid line). {The dashed lines of respective colours show the inputs from photoionisation heating only in models\,\#1 and \#2.} The two black lines at the top show the relative cooling input for model\,\#1 (dashed-dotted line) and model\,\#2 (solid line). The grey solid (model\,\#2) and dashed-dotted (model\,\#1) lines show for comparison the heating efficiency given by the approximation of \citet{mc2009} for the wavelengths that provide the largest input at the given radial distance. In each panel, the two vertical lines give the position of the two maxima of the volume heating rate.}
    \label{fig::6.2b_HE}
\end{figure}%
\begin{figure}
    \centering
    \includegraphics[width=0.97\columnwidth]{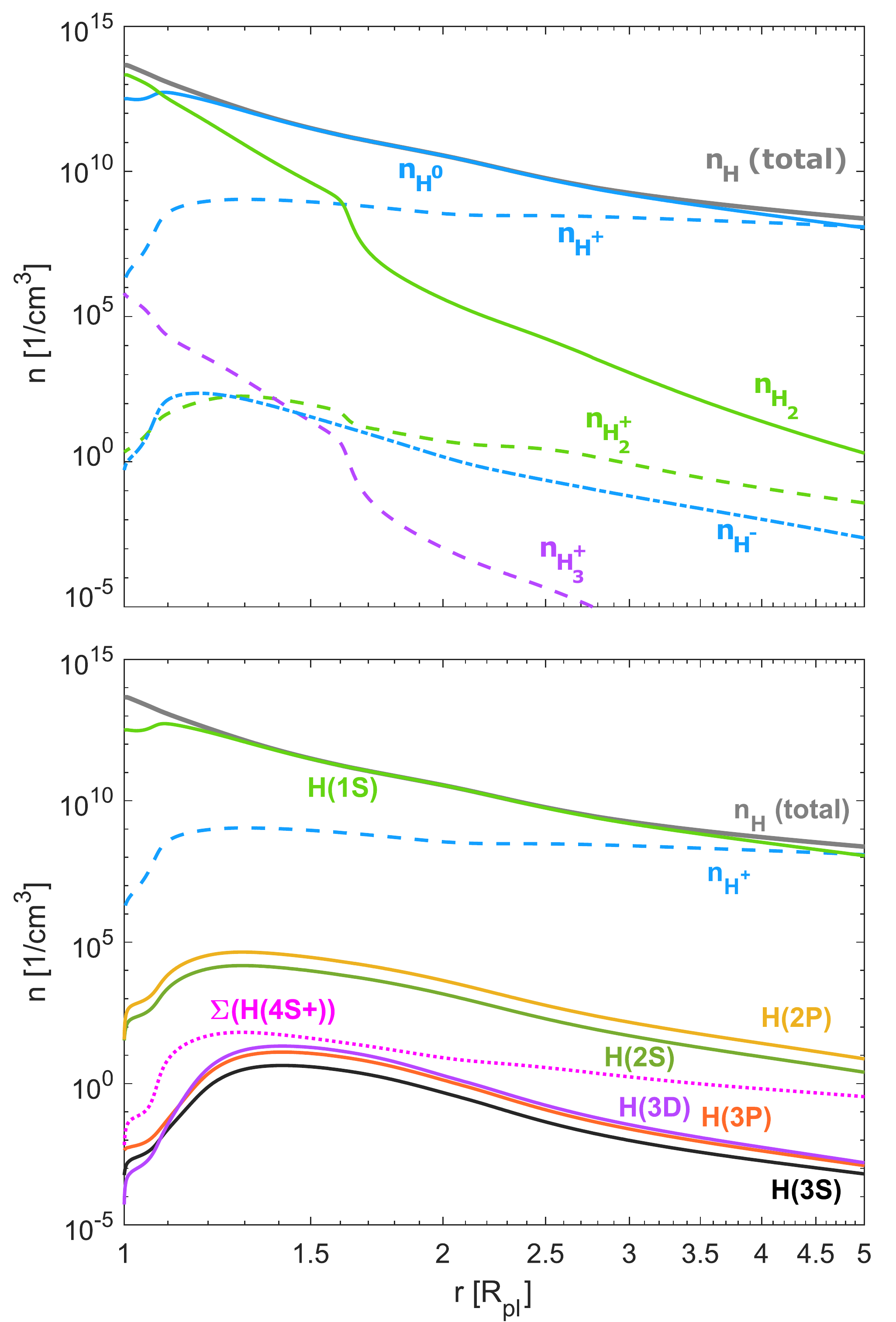}
    \caption{{Distribution of hydrogen species against radial distance predicted by model\#1. Top panel: numerical densities of neutral atomic hydrogen (blue solid line), its ions (blue dashed), and ${\rm H^-}$ (blue dashed-dotted), molecular hydrogen (green solid) and its ions (green dashed), and ${\rm H_3^+}$ (violet dashed). Bottom panel: distribution of neutral atomic hydrogen in terms of excitation levels, i.e. numerical densities of hydrogen in 1S (light green solid line), 2S (dark greed solid), 2P (dark yellow solid), 3S (black solid), 3P (orange solid), and 3D (violet solid) states. The dotted magenta line gives the summed numerical densities of hydrogen atoms in the energy levels 4S and higher. In both panels, the grey solid line gives the total density of all hydrogen species for reference.}}
    \label{fig::6_2b_Hpop}
\end{figure}
{The differences described above are due to differences in the treatment of how the atmosphere absorbs the stellar radiation.} The profiles of the volume heating rate obtained including Cloudy (models\,\#1 to \#5) {have considerably different shapes compared to those} obtained from the hydrodynamic model alone (model\,\#0), but are still characterised by two peaks {in the region dominated by the absorption of X-ray (i.e. $\lambda\lesssim10$\,nm; lower in the atmosphere) and EUV (i.e. $\lambda\sim10-90$\,nm; higher in the atmosphere)} radiation, though spatially more separated, with the X-ray-dominated peak occurring closer-in and the EUV-dominated lying further away from the planet. Panel (a) of Figure\,\ref{fig::6.2b_HE} shows the input from the specific wavelengths into the total ionising flux as a function of atmospheric radial distance. Interestingly, the peak in the lower part of the atmosphere ({associated with} absorption of X-ray photons) predicted by models\,\#1--\#5 is {${\sim1.5-4.5}$ times} higher than for model\,\#0, while the other peak ({associated with} absorption of EUV photons) appears to be considerably lower. The latter results from the difference in atmospheric density of {2.1--4.4 times} between model\,\#0 and models \#1--\#2 at the position of the EUV-related peak near $\sim2$--2.5\,\Rpl\ (see panel (b) of Figure\,\ref{fig::6_2b_profiles}). This can be verified by comparing the volume heating function of model\,\#0 with the one calculated following the first iteration involving Cloudy within model\,\#1, i.e. the two heating functions calculated using different approaches, but for the same density and temperature profiles (see Figure\,\ref{fig::6.2b_1st}). Here, the peaks due to absorption of EUV photons predicted by models\,\#0 and \#1 have similar heights, while the peaks connected to the absorption of X-ray photons have different amplitudes. Therefore, the strong difference in EUV-related peaks comes, mainly, not from the differences in photoionisation models but from the change in atmospheric parameters (over thousands of code iterations) due to the increase of the heating in the lowermost atmospheric layers, i.e. the reduction of the EUV peak is a direct consequence of the increase of the X-ray-related peak.

The reason for the increased heating in the X-ray-dominated region relates to one of the simplifications of the hydrodynamic code (model\,\#0), which assumes that the whole heating is produced by photoionisation with a fixed heating efficiency of 0.15 for both X-ray and EUV flux (parameter $\eta$ in Equation\,(\ref{eq::Qxuv})) that is constant with atmospheric height. \citet{salz2016} showed that this simplification is adequate for planets with gravitational potentials up to $\sim10^{13.1}$\,erg\,g$^{-1}$, at least for what concerns estimating mass-loss rates. However, the realistic shape of the heating efficiency (i.e. of the fraction of the total absorbed energy spent on heating; see Section~\ref{ssec::HE_note}) is more complicated as $\eta$ depends on the wavelength of the incoming stellar radiation and atmospheric depth \citep[e.g.][]{waite1983,dalgarno1999,yelle2008,shematovich2014,ionov_shematovich2015,GMunoz2023A&A...672A..77G}. %

Differently from the basic hydrodynamic model, Cloudy solves the local equilibrium state in each cell of the simulation domain. It solves not only the ionisation/recombination and chemical reactions for each species, but in doing this it also accounts for the level populations of the single energy levels, and thus it implicitly computes the heating efficiency self-consistently. {Figure\,\ref{fig::6_2b_Hpop} shows the height distribution of different hydrogen species, including neutrals and ions of hydrogen atoms and molecules (top panel) and the neutral atomic hydrogen in different excitation states (bottom panel) according to the predictions of model\,\#1. One can see that the narrow lowermost region of the atmosphere ($r\lesssim1.1$\,$R_{\rm pl}$) is dominated by neutral molecular hydrogen, while the region from $1.1-4.5$\,$R_{\rm pl}$, where most of the heating takes place, is dominated by neutral atomic hydrogen. Above this region (above the temperature maximum, where the atmosphere becomes optically thin at all XUV wavelengths), the main constituent is hydrogen ions. Furthermore, the ion distribution peaks at about 1.25\,$R_{\rm pl}$, i.e. the position of the X-ray-dominated maximum of the heating function. The number of heavier ions ${\rm H_2^+}$ and ${\rm H_3^+}$, as well as ${\rm H^-}$ maximises below 1.5\,$R_{\rm pl}$ and, in general, remains a few orders of magnitude smaller than for the species discussed above. In terms of energy levels of atomic hydrogen (whose ionisation contributes most to the heating of the atmosphere), the ground state (1S) is the most populated level, followed by 1P and 2S states, though $\sim7$ orders of magnitude less abundant. The bottom panel of Figure\,\ref{fig::6_2b_Hpop} shows the distribution for the energy levels up to 3D, and the distribution of atoms with higher energies is given as their sum (dotted magenta line); in total, Cloudy accounts for 69 hydrogen energy levels.}

Given that Cloudy accounts for all considered opacity sources, the code also accurately computes the fraction of stellar flux that is absorbed at each specific height in the atmosphere (panel (b) of Figure\,\ref{fig::6.2b_HE}). Panel (c) of Figure\,\ref{fig::6.2b_HE} shows the heating efficiency calculated a-posteriori from Cloudy outputs across the planetary atmosphere as the ratio between the heating rate and the total XUV deposition rate 
\begin{eqnarray}\label{eq::HE}
    \eta_{\rm chy} &=& \frac{H_{\rm tot}(r)}{W_{\rm XUV}(r)} \nonumber \\
    &=& {H_{\rm tot}}\left( \sum_{\rm i}4\pi n_{\rm i}\int_{0}^{\lambda_{\rm i,0}}{J_{\lambda}}\sigma_{\rm \lambda,i}\exp(-\tau(\lambda, r))d\lambda \right)^{-1}\,.
\end{eqnarray}
Here, $n_{\rm i}(r)$ is the numerical density of species ``i'' relative to the hydrogen density, {$J_{\lambda}$ is the stellar flux at the specific wavelength $\lambda$ in erg\,s$^{-1}$\,cm$^{-2}$, $\sigma_{\rm \lambda,i}$ is the absorption cross-section of species ``i'' at the corresponding wavelength}, and $\tau$ is the optical depth. The wavelength $\lambda_{\rm i,0}$ corresponds to the minimum photon energy necessary to ionise species ``i''. As separate species, we consider here the different chemical elements and their ionised and excited states. For model\,\#1, the heating efficiency reaches its maximum below 1.5\,\Rpl, close to the position corresponding to the maximum of the volume heating rate {associated with} the absorption of X-ray photons. Above this point, the fraction of the absorbed radiation that is spent on heating decreases, until it reaches a constant value of about 0.4 (see the blue dashed-dotted line in panel (c) of Figure\,\ref{fig::6.2b_HE}). Therefore, within model\,\#1 the heating efficiency at the position of the maximum of the volume heating rate due to absorption of EUV photons ($\sim2.3$\,\Rpl) is about half of that corresponding to the maximum {associated with} absorption of X-ray photons. {However, the total heating rate $H_{\rm tot}$ in Equation\,(\ref{eq::HE}) includes the input from processes other than photoionisation. Specifically, the prominence of the left peak in the volume heating function and the maximum in heating efficiency in model\#1 are associated with the iso-sequence line heating term $H_{\rm lin}$ (see Section\,\ref{sec::heat-cool} for detail). For the considered case, the efficiency of photoionisation itself maximises at photon energies of $\sim$25--60\,eV with a peak at $\sim$32\,eV. The higher-energy photoelectrons are expected to lose their energy in the cascade ionisation and internal excitation collisions with atomic hydrogen rather than in elastic collisions, which reduces their contribution to the heating of the atmosphere \citep[e.g.][]{ionov_shematovich2015}. For comparison, we include in Figure\,\ref{fig::6.2b_HE} the efficiencies associated with photoionisation of atomic and molecular hydrogen only (i.e. $H_{\rm tot}$ in Equation\,(\ref{eq::HE}) is substituted with $H_{\rm{I}}+H_{\rm 2ph}$; see Section\,\ref{sec::heat-cool} for detail).}

With the inclusion of wind advection (model\,\#2), heat is transported upwards {and the $H_{\rm lin}$ term is significantly suppressed. Therefore, $\eta$ at the peak associated with the absorption of X-ray photons decreases,} while it does not vary in the upper atmospheric layers (see the magenta solid line in panel (c) of Figure\,\ref{fig::6.2b_HE}). In turn, this also leads to modifications in the cooling fraction unaccounted for in the calculation of the heating efficiency (black dashed-dotted and solid lines corresponding to models\,\#1 and \#2, respectively, in panel (c) of Figure\,\ref{fig::6.2b_HE}) where the maximum cooling input decreases in model\,\#2 as a result of upwards heat transport.

In terms of the atmospheric parameters, the increase in the heating below $r = 1.5$\,\Rpl\ in models\,\#1 and \#2 compared to that in model\,\#0 leads to the emergence of a planetary wind in deeper and denser atmospheric layers. Due to the more effective expansion, the density of the outflow above $\sim2$\,\Rpl\ increases, and the sonic point is pushed outwards from ${ \sim2.7}$\,\Rpl\ to $\sim3.5$\,\Rpl. The inclusion of wind advection leads to upward redistribution of the heat, and thus the X-ray heating that in model\,\#1 causes the secondary temperature maximum at $\sim1.5$\,\Rpl\ (see the light-blue dashed-dotted line in panel (c) of Figure\,\ref{fig::6_2b_profiles}), in model\,\#2 can only lead to a less pronounced temperature minimum in comparison to model\#0 (see magenta solid line compared to dark-blue dotted line).
\subsubsection{A note on heating efficiency}\label{ssec::HE_note}
The definition of heating efficiency is not unique across the literature and therefore direct comparisons might not be possible. In a most general sense, the heating efficiency relates the fraction of incoming (stellar) energy converted into heat with the total energy absorbed by the atmosphere. This value depends on the wavelength of the incoming radiation, on the atmospheric depth, and on the species absorbing the radiation (see Equation\,(\ref{eq::HE})). However, while the value of the total XUV deposition rate ($W_{\rm XUV}$) is clearly defined, the heating rate ($H_{\rm tot} - C_{\rm tot}$) depends on the considered physical processes and on their specific treatment. The most common approximation is that the heating is driven mostly by photoionisation (as in model\,\#0) and the whole excess photon energy is converted into heating the atmosphere. In this case, the heating rate present in Equation\,(\ref{eq::HE}) can be expressed as
\begin{equation}\label{eq::H_PE}
    H_{\rm PE} = \sum_{\rm i}4\pi n_{\rm i}\int_{0}^{\lambda_{\rm i,0}}{J_{\lambda}}\frac{(E_{\lambda} - E_{\rm i})}{E_{\lambda}}\sigma_{\rm \lambda,i}\exp(-\tau(\lambda, r))d\lambda\,,
\end{equation}
where $E_{\rm i}$ is the (threshold) ionisation energy of species ``i'' {and $E_{\lambda}$ is the photon energy}. By considering a hydrogen atmosphere and the $H_{\rm PE}/W_{\rm XUV}$ ratio, the maximum heating efficiency can be approximated as $\eta_{\rm max} = 1 - 13.6{\rm eV}/E_{\lambda}$ \citep{mc2009,waite1983}. For comparison with the value of $\eta$ obtained using Equation\,\ref{eq::HE}, panel (c) of Figure\,\ref{fig::6.2b_HE} (grey dashed-dotted and solid lines) also shows $\eta_{\rm max}$ computed considering photoionisation of atomic hydrogen and just the wavelengths of the stellar irradiation that give the maximum yield in heating at a specific radial distance, as illustrated in panel (a) of Figure\,\ref{fig::6.2b_HE}.
\begin{figure}
    \centering
    \includegraphics[width=0.97\columnwidth]{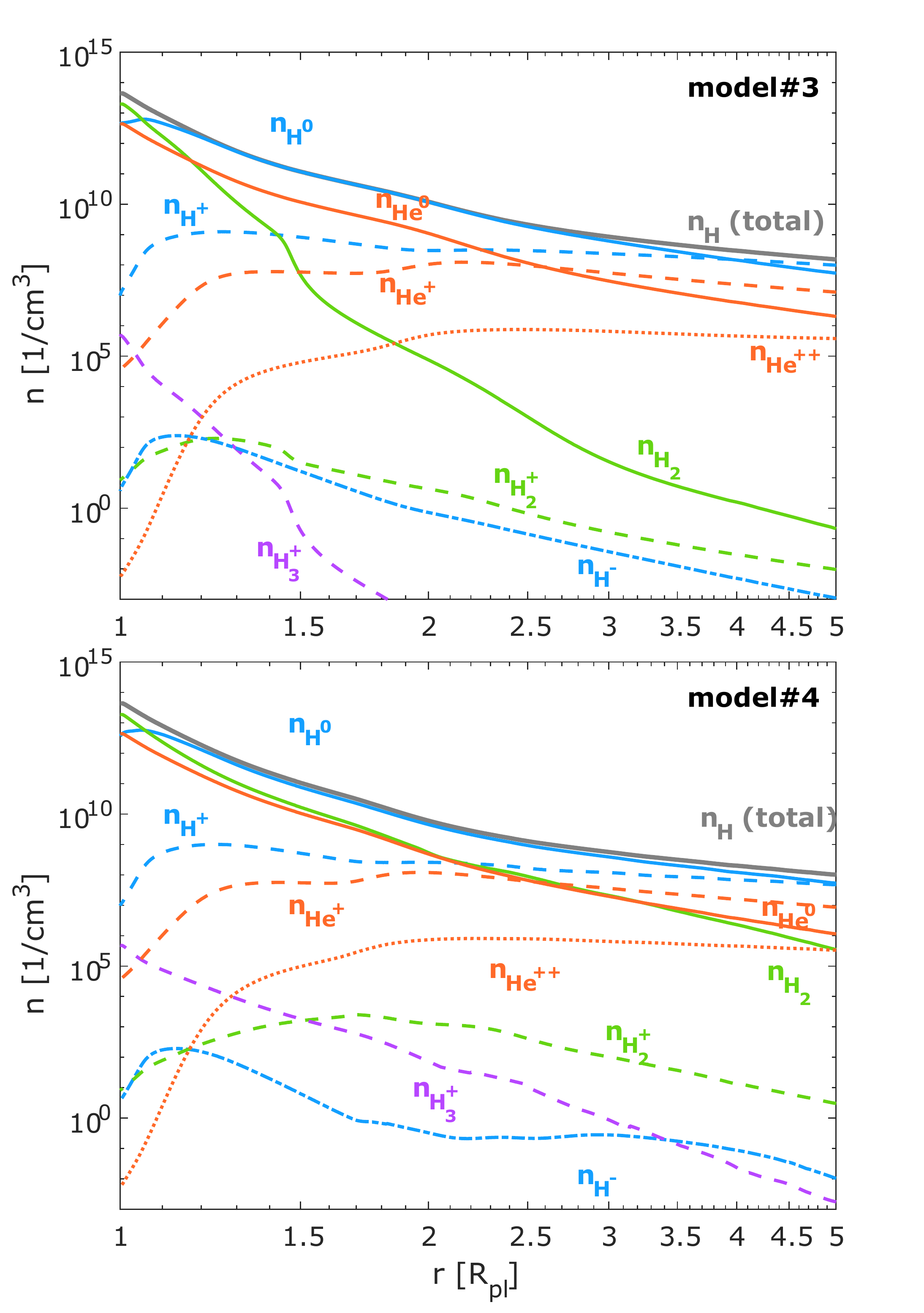}
    \caption{{Distribution of hydrogen and helium species against radial distance predicted by models\,\#3 (top panel) and \#4 (bottom panel). The grey solid line gives the total numerical density of all hydrogen species, for reference. The blue lines show the species of atomic hydrogen (solid for neutral H, dashed for ${\rm H^+}$, and dashed-dotted for ${\rm H^-}$), the green lines show molecular hydrogen (solid for neutral ${\rm H_2}$ and dashed for ${\rm H_2^+}$), the violet dashed line shows ${\rm H_3^+}$, and the orange lines show helium species (solid for neutral He, dashed for ${\rm He^+}$, and dotted for ${\rm He^++}$).}}
    \label{fig::6_2b_He_cond}
\end{figure}

In general, the approximation given by Equation\,(\ref{eq::H_PE}) leads to overestimating the heating efficiency \citep{shematovich2014}. In reality, the excess energy released in the photoionisation reaction is not transferred into heat directly, but passed instead to the products of the photoionisation reaction, i.e., mainly to the free electron. The energetic electron can further lose this energy in elastic and inelastic collisions with other particles, where the latter can lead to secondary ionisation or excitation of atmospheric species. %
Furthermore, as we will show in the next section, heating (and cooling) processes are not limited to photoionisation.

Several analytical models and hydrodynamic codes consider a constant heating efficiency that is not computed self-consistently. These parameters are normally estimated (semi-)empirically to predict as realistically as possible atmospheric mass-loss rates and are expected to parameterise a wide range of physical processes. For example, inserting the value of $\eta = 0.15$ employed in model\#0 in the analytical energy-limited approximation \citep{watson1981} returns mass-loss rates in line with those predicted by \citet{salz2015}, which used a code comparable to what presented here. Alternatively, \citet{owen_wu2017} suggested a value of $\eta = 0.1(v_{\rm esc}/15\,{\rm km/s})^{-2}$, where $v_{\rm esc}$ is the escape velocity at the photosphere. This value of $\eta$ has been chosen, because it leads one to reproduce the radius valley \citep{fulton2017} employing the energy-limited approximation. Therefore, these approximations of the heating efficiency should not be directly compared with the values obtained self-consistently.
\subsubsection{Hydrogen-helium atmospheres and metal heating}\label{sec::heat-cool}
With the inclusion of helium (models\,\#3 and \#4), the atmospheric mass-loss rates decrease by a factor of $\sim1.3$ without accounting for wind advection and of $\sim1.15$ with wind advection {compared to the pure hydrogen case}. Correspondingly, the outflow density decreases, respectively, by factors of 1.5 and 1.3, and the outflow velocity increases by about a factor of 1.15. Independently of the inclusion of wind advection, the maximum temperature increases $\sim1.3$ times (see Table\,\ref{tab::6_2B_cmodels}), 
which might suggest a decrease in the atmospheric heating in the lowermost layers of the atmosphere. However, an inspection of the temperature profiles in Figure\,\ref{fig::6_2b_profiles} indicates the opposite, as the temperature maximum corresponding to X-ray heating is higher than that obtained for a pure hydrogen atmosphere. Therefore, the variations in the outflow parameters described above are caused mainly by the increase in mean molecular weight.

{In terms of the distribution of different atmospheric species, the picture is similar to the pure hydrogen case (see Figure\,\ref{fig::6_2b_He_cond} for {models\,\#3 and \#4). For model\,\#3, the profiles} of hydrogen species look nearly the same as in model\,\#1, though all the characteristic features move slightly inwards. The total fraction of helium (relative to the total hydrogen density) is kept constant across the simulation domain; this might not always be the case in real atmospheres, but accounting for such effects requires employing multifluid hydrodynamics, as we noted above. Therefore, near the lower boundary neutral helium is a minor constituent relative to both atomic and molecular neutral hydrogen; with the steep decrease of the ${\rm H_2}$ abundance and the increase in ion fraction, however, neutral helium becomes the second most abundant element between 1.2 and 2.5\,$R_{\rm pl}$, where most of the heating takes place. The number density of helium ions, ${\rm He^+}$ and ${{\rm He^{++}}}$, is maximum near the EUV-dominated peak of the heating function, in contrast to the behaviour of the hydrogen ions; therefore, ionisation of helium contributes to EUV heating rather than to X-ray heating. The ionisation barrier (the point where the number of ions overcomes that of the neutral atoms for the specific species) for helium occurs closer to the planet compared to hydrogen ($\sim2.5$\,$R_{\rm pl}$ against $\sim4.0$\,$R_{\rm pl}$). This occurs because the number of ions in this region depends on the XUV irradiation level rather than on the available reservoir of neutral atoms: thus, $n_{\rm He} = 0.1n_{\rm H}$ but $n_{\rm He^+} = \sim0.3n_{\rm H^+}$ near the helium ionisation barrier, meaning that the relation of the number density of ions to that of neutrals is higher for helium than for hydrogen there.}
{The inclusion of wind advection in model\,\#4 (bottom panel of Figure\,\ref{fig::6_2b_He_cond}) affects the distribution of the lighter hydrogen species rather than that of helium species. Specifically, hydrogen species are dragged upwards by the hydrodynamic wind, while their relative occurrence in the lowermost layer remains unaltered. The effect on neutral and ionised atomic hydrogen (see $n_{\rm H^0}$ and $n_{\rm H^+}$) is limited, and they remain the dominant constituents in the upper atmospheric layers. The fraction of molecular hydrogen species (see $n_{\rm H_2}$, $n_{\rm H_2^+}$, and $n_{\rm H_3^+}$) in these upper layers, however, increases significantly, though it remains minor compared to that of the atomic hydrogen species.}

In the following, we consider in more detail the specific processes that contribute to atmospheric heating in the frame of models\,\#1, \#3, and \#5. For simplicity, we consider the runs not accounting for wind advection as the relative input from the different processes does not change significantly with the inclusion of wind advection, and the difference between models\,\#1 and \#2, and \#3 and \#4 is mainly due to the increased advection term {(which dominates cooling above $\sim1.5-2$\,$R_{\rm pl}$)}, while the radiative components remain similar.

\begin{figure}
    \centering
    \includegraphics[width=0.97\columnwidth]{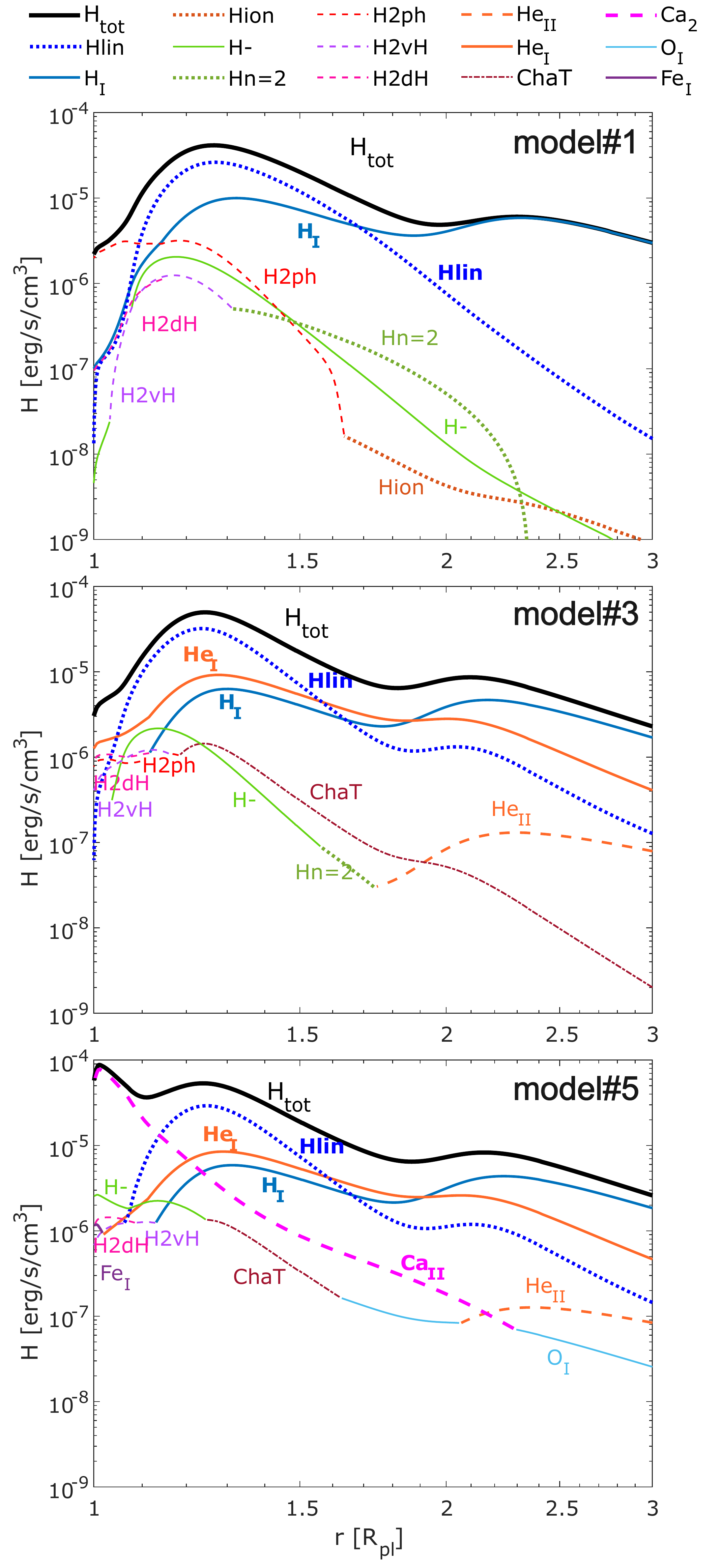}
    \caption{Volume heating rates and their most relevant contributors ($>$5\% of $H_{\rm tot}$ at each specific $r$) calculated for the hot 5\,\Mer\ sub-Neptune under moderate XUV irradiation. The top, middle, and bottom panels present the results of model\,\#1 (pure hydrogen atmosphere), model\,\#3 (hydrogen$+$helium), and model\,\#5 (solar composition), respectively. The processes responsible for the heating and the corresponding line styles are given in the legend located above the top panel and follow this notation: ${\rm H_{tot}}$ -- total volume heating rate from all sources; Hlin -- iso-sequence line heating; H{\sc i} -- heating due to photoionisation of atomic hydrogen; Hion -- collisional ionisation heating (all hydrogen species); H- -- H$^-$-heating; ${\rm Hn=2}$ -- photoionisation from all excited states of hydrogen species; H2ph -- H$_2$ photoionisation heating; H2vH -- heating by collisional de-excitation of vibrationally excited H$_2$; H2dH -- heating by photodissociation of H$_2$; He{\sc ii} -- photoionisation of He{\sc ii}; He{\sc i} -- photoionisation of He{\sc i}; ChaT -- charge transfer heating; Ca{\sc ii} -- metal line heating dominated by absorption of the Ca{\sc ii} line at 7291.47\,\AA; O{\sc i} -- photoionisation of O{\sc i}; Fe{\sc i} -- photoionisation of Fe{\sc i}.}
    \label{fig::6_2b_heating}
\end{figure}

In Figure\,\ref{fig::6_2b_heating}, the black solid lines show the total volume heating rate $H_{\rm tot}$, while the other lines indicate specific heating processes, as given by the legend. These processes are only shown in the regions, where they contribute not less than 5\% to the total heating rate. %
For all three models, the peak at about 1.2\,\Rpl\ of the heating function is dominated by the iso-sequence line heating (Hlin, dark-blue dotted line in Figure\,\ref{fig::6_2b_heating}), that is line heating due to all hydrogen species, which declines with decreasing density and contributes no more than $\sim10\%$ of the total heating beyond $r\sim2\Rpl$. {Given that heating in the lowermost atmospheric levels has a minor impact on the general outcome of the model (see the discussion in Section\,\ref{sec::code_bord}), we tested the impact of the Hlin term on our predictions by re-running the code for a few planets in the grid, explicitly excluding the Hlin heating. We found that, despite its relative prominence, Hlin contribution is negligible in terms of atmospheric mass-loss rates and only slightly affects the atmospheric temperature and ion fractions.} The outermost peak of the volume heating profile is instead dominated by ionisation of atomic hydrogen (${\rm H_1}$; solid blue line in Figure\,\ref{fig::6_2b_heating}), and at short radial distances, where Hlin dominates, this process contributes up to 25\% of the total heating rate. 

In the case of model\,\#1 (atmosphere consisting solely of hydrogen species), below $r\leq1.1$\,\Rpl\ the heating is mainly controlled by the ionisation of molecular hydrogen. However, the abundance of ${\rm H_2}$ steeply declines with height due to photodissociation, and thus H$_2$ ionisation does not play a significant role above $\sim1.3$\,\Rpl. Other processes associated with hydrogen species contribute just a minor part to the total heating rate.

With the inclusion of helium in model\,\#3, the heating increases by $\sim25-50\%$ in the lower part of the atmosphere below $\sim2.5$\,\Rpl, predominantly due to the ionisation of atomic helium by photons more energetic than 24.5\,eV (He{\sc i}; solid orange line in Figure\,\ref{fig::6_2b_heating}). This process also overcomes the photoionisation of molecular hydrogen at the bottom of the atmosphere {(though the densities of both neutral and ionised helium are lower than those of hydrogen in this region; see Figure\,\ref{fig::6_2b_He_cond})}. The ionisation of He{\sc ii}, in turn, contributes to the heating about an order of magnitude less than the ionisation of neutral atomic helium and is only accountable above $\sim1.75-2$\,\Rpl. Furthermore, with an increase of the temperature gradient, charge transfer heating increases at $\sim1.2-3$\,\Rpl, but it remains, however, a minor contributor.

\begin{figure}
    \centering
    \includegraphics[width=0.97\columnwidth]{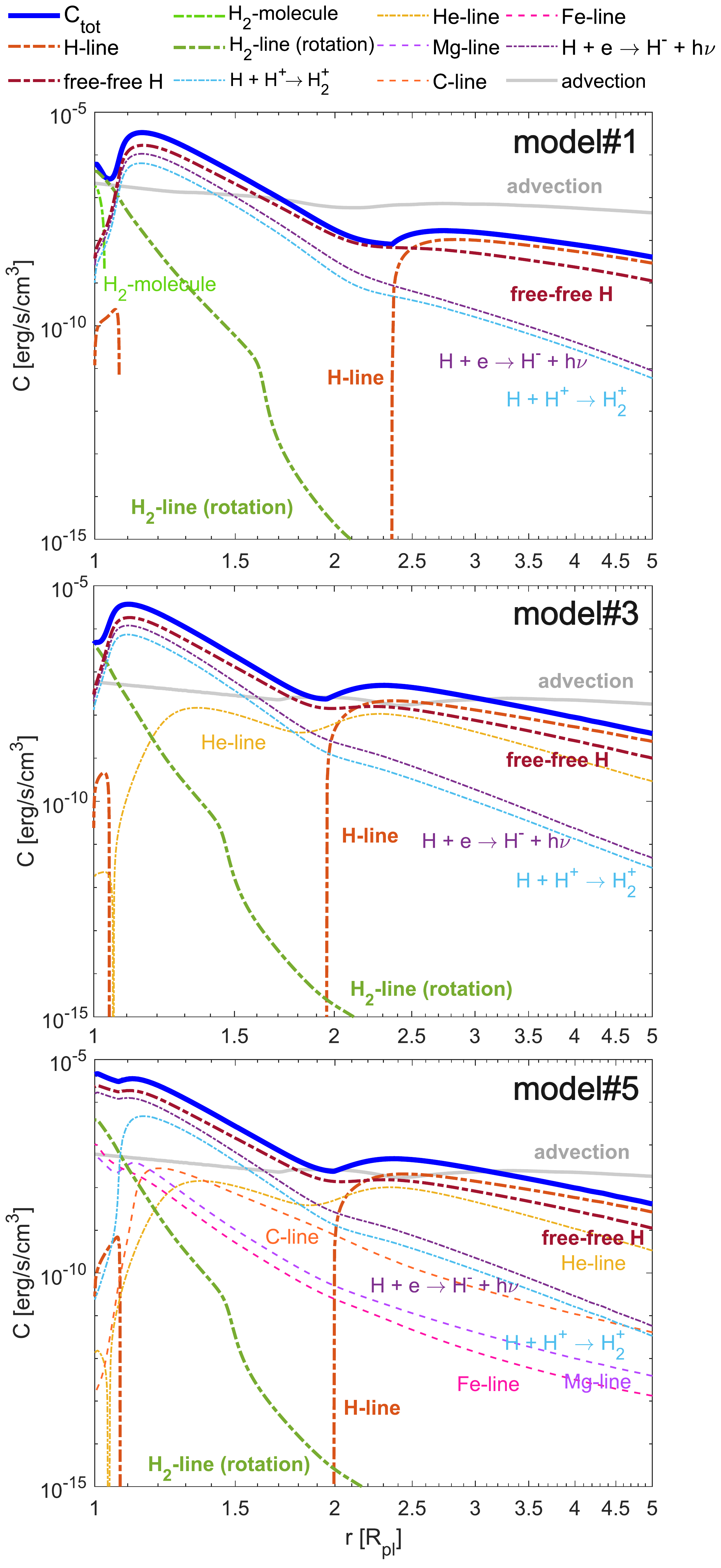}
    \caption{Volume cooling rates and their most relevant contributors ($>$1\% of $C_{\rm tot}$ at their maximum) calculated for the hot 5\,\Mer\ sub-Neptune under moderate XUV irradiation. The top, middle, and bottom panels present the results of model\,\#1 (pure hydrogen atmosphere), model\,\#3 (hydrogen$+$helium), and model\,\#5 (solar composition), respectively. The processes responsible for the cooling, and the corresponding line styles are given in the legend located above the top panel and follow this notation: ${\rm C_{tot}}$ -- total volume cooling from all sources; H-line -- atomic hydrogen line cooling; free-free H -- bremsstrahlung cooling from hydrogen (and helium, when included in the model); ${\rm H_2}$-molecule -- ${\rm H_2}$ molecule cooling; ${\rm H_2}$-line (rotation) -- line cooling within a ${\rm H_2}$ molecule; {${\rm H + H^+\rightarrow H_2^+}$ -- cooling due to recombination of ${\rm H_2^+}$}; ${\rm H + e \rightarrow H^- + h\nu}$ -- cooling due to an electron hitting a hydrogen atom in the ground state loosing its kinetic energy and hydrogen atom can emit a photon; He-line -- helium line cooling; C-line -- carbon line cooling; Mg-line -- magnesium line cooling; Fe-line -- iron line cooling. For reference, we include in each panel the advection term including adiabatic cooling from the models with the wind, with their respective composition.}
    \label{fig::6_2b_cooling}
\end{figure}

Finally, in model\,\#5 which assumes solar atmospheric abundances, the heating sources discussed above remain at a similar level. Additionally, some of the metal heating processes contribute considerably to the heating function. Among them, the major input comes from absorption by the Ca{\sc ii} line at 7291.47\,\AA, which pushes the heating at $r\leq1.1$\,\Rpl\ significantly above the value predicted by model\,\#3 (i.e. without metals). However, for the considered planet, this does not have significant consequences for the atmospheric temperature or outflow parameters. In general, this heating source has a noticeable impact just for the hottest planets under low-to-moderate XUV irradiation and we discuss this in more detail in Section\,\ref{sec::gridtests_muhe}. {We note that in this specific model the ${\rm\mu}$bar pressure level is located at about 1.2\,$R_{\rm pl}$. Given that the temperature in this region is close to 1500\,K, condensation might lead to a lack of Ca in the upper atmosphere, thus making the impact of this specific heating term insignificant.}

The total cooling rates change between models \#1, \#3, and \#5 similarly to the heating rates discussed above (see Figure\,\ref{fig::6_2b_cooling}). Namely, the total cooling rate increases slightly in the lowermost part of the atmosphere (below $\sim1.2-1.3$\,\Rpl) from model\,\#1 to model\,\#5. However, the cooling rates remain at least an order of magnitude lower compared to the heating rates, except at $r\leq1.1$\,\Rpl\ for the hydrogen-only and hydrogen-helium atmospheres, where heating and cooling rates have similar values. {In the models\,\#2 and \#4 accounting for wind advection, cooling in the outer regions (above 1.5--2.0\,$R_{\rm pl}$) is dominated by cooling due to adiabatic expansion. In Cloudy, this term is hidden in the advection cooling term, which we include in the plots for reference.}

Most of the cooling in all three models is produced by line cooling of atomic hydrogen (H-line; light-brown line in Figure\,\ref{fig::6_2b_cooling}) and bremsstrahlung cooling from hydrogen and helium (free-free H; dark-brown line). At $r\leq2$\,\Rpl, up to 30\% of the cooling is provided by {the production of ionised molecular hydrogen (${\rm H + H^+ \rightarrow H_2^+}$; light-blue line)} and collisions of neutral hydrogen atoms with energetic electrons (i.e. secondary ionisation; ${\rm H + e \rightarrow H^- + h\nu}$; purple line). Molecular hydrogen line cooling (dark-green line) dominates the cooling at $r\leq1.1$\,\Rpl\ for models\,\#1 and \#3. Furthermore, in models\,\#3 and \#5, helium line cooling (yellow line) contributes up to 20\% of the total cooling rate close to the position of the maximum of the heating function corresponding to the absorption of the EUV radiation. However, the total cooling rate in this region of the atmosphere is 2--3 orders of magnitude smaller than heating. Finally, metal line cooling rates (see C-line, Mg-line, and Fe-line in the bottom panel of Figure\,\ref{fig::6_2b_cooling}) contribute no more than $\sim1\%$ of the total cooling rate.
\subsection{Impact of accurate energy balance computation for Neptune-like planets with hydrogen atmospheres}\label{sec::gridtests_grid}
\begin{figure*}
    \centering
    \includegraphics[width=0.8\hsize]{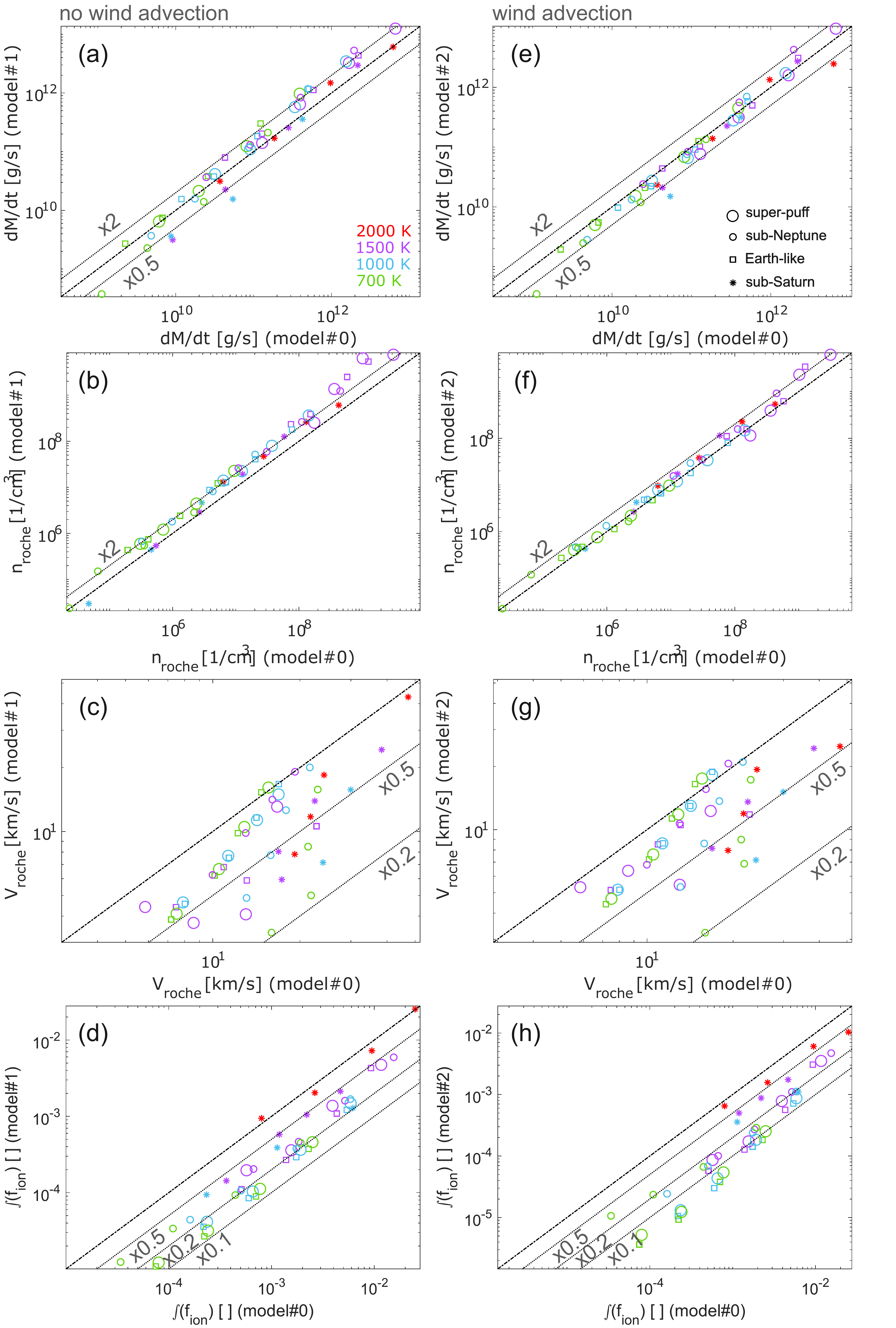}
    \caption{Comparison of the basic parameters of the atmospheric outflow between models\,\#0 and \#1 (panels (a)--(d)) and models\,\#0 and \#2 (panels (e)--(h)). Specifically, we compare the atmospheric mass-loss rate (panels (a) and (e)), the outflow density at $R_{\rm roche}$ (panels (b) and (f)), {the outflow velocity at $R_{\rm roche}$} (panels (c) and (g)), and the total fraction of ions below $R_{\rm roche}$ (panels (d) and (h)). In each panel, the x-axis corresponds to the value of the specific parameter predicted by model\,\#0 and the y-axis corresponds to the same parameter predicted by model\,\#1 (panels (a)--(d)) or model\,\#2 (panels (e)--(h)). Different colours correspond to different equilibrium temperatures and different symbols correspond to different planet types, as indicated in the legends in panels (a) and (b). The black dashed-dotted lines in each panel show the equality of the parameters predicted by the two models, and the additional grey lines show different ratios between the parameters $P$ obtained from the different models in the form $P_{\rm model\,\#1/\#2} = c \times P_{\rm model\,\#0}$, where the coefficient $c$ is shown in the plots next to each grey line.}
    \label{fig::comp_hydro_H}
\end{figure*}

In this section, we overview the effects of the inclusion of the precise energy balance computation provided by Cloudy on the atmospheric outflow from low-mass planets. Figure\,\ref{fig::comp_hydro_H} shows the comparison among some basic parameters predicted by the hydrodynamic model\,\#0 to those predicted by model\,\#1 (i.e. without wind advection; left column) and by model\,\#2 (with wind advection; right column). The differences between the results obtained for model\,\#0 and \#1 and for model\,\#0 and \#2 are qualitatively similar, therefore we start with comparing the results of models\,\#0 and \#1 and then discuss the impact of the inclusion of wind advection (model\,\#2).

One of the most important parameters of atmospheric outflow, particularly relevant for planetary evolution, is the mass-loss rate. Panel (a) of Figure\,\ref{fig::comp_hydro_H} shows that the values predicted by model\,\#1 are in {most cases} slightly higher than those predicted by model\,\#0. {In general, the difference does not exceed about a factor of two (see the grey line in panel (a)), and the increase in escape is maximum at high irradiation levels and for low planetary masses. In contrast, for planets with high mass and/or low temperature, the mass-loss rates predicted by model\#1 are lower than those predicted by model\#0 (e.g. planets 1.2B, 1.2C, 4.4B, and 4.4C in Table\,\ref{tab::gridplanets}). This trend remains if one considers a planet with a fixed mass and temperature. This suggests that the increase in heating efficiency of high-energy photons discussed in Section\,\ref{sec::gridtests_HE} (and the X-ray heating itself) is most relevant for young, highly irradiated low-mass planets, where the ``X-ray dominated'' region of the atmosphere can be relatively extended. Instead, the accurate treatment of cooling processes is most relevant for weakly irradiated planets, where they compete with photoionisation heating.}

With the inclusion of wind advection (model\,\#2), the differences {with model\,\#0 decrease slightly and the turning point between high and low XUV regimes shifts to higher mass-loss rates (from $\sim10^{10}$ to $\sim10^{11}$\,g/s; see panels (a) and (e) of Figure\,\ref{fig::comp_hydro_H}) due to a more consistent inclusion of adiabatic cooling.}
The correlations with planetary mass, temperature, and XUV flux, however, remain the same. Therefore, for both models\,\#1 and \#2, the differences in mass-loss rates compared to model\,\#0 are quite low and are expected to be of little relevance in terms of atmospheric evolution.

At the Roche radius, the behaviour of the atmospheric density is similar to that of the mass-loss rate (see panels (b) and (f) of Figure\,\ref{fig::comp_hydro_H}). Model\,\#1 predicts about {two--four} times higher outflow densities compared to model\,\#0, but its dependence on planetary mass and incident XUV flux is not as pronounced as for the mass-loss rate, while the dependence on equilibrium temperature is comparable. {With the inclusion of wind advection, the difference from model\,\#0 nearly vanishes, except for low-mass highly irradiated planets. We also find that for weakly irradiated planets the decrease in mass-loss rate comes predominantly from the decrease in outflow velocity.}

{The relation between the outflow velocity $V_{\rm roche}$ predicted by models\,\#0 and \#1 (see panel (c) of Figure\,\ref{fig::comp_hydro_H}) is steeper and more spread than those discussed above. It is similar to the relation of maximum temperature $T_{\rm max}$, which we thus do not show; furthermore, as one can see from Figure\,\ref{fig::6_2b_profiles}, the peak temperature values do not reflect well enough all the variations in the temperature profiles between different models, which are more complicated than a simple shift at lower/higher temperatures. }

{The total range of outflow velocities does not change much between the two models, with lower and upper limits shifting by only $\sim3$\,km\,s$^{-1}$ towards lower values in model\,\#1. However, the changes for individual planets can reach a factor of five and are maximum for cooler and more massive planets. The dependence of $V_{\rm roche}$ on the stellar XUV flux is similar to that of mass-loss rate and density, though steeper.} 
{The inclusion of wind advection in model\,\#2 does not alter the difference to model\#0 in terms of velocity qualitatively, but softens the dependence of $V_{\rm roche}$ on XUV.}

\begin{figure}
    \centering
    \includegraphics[width=\columnwidth]{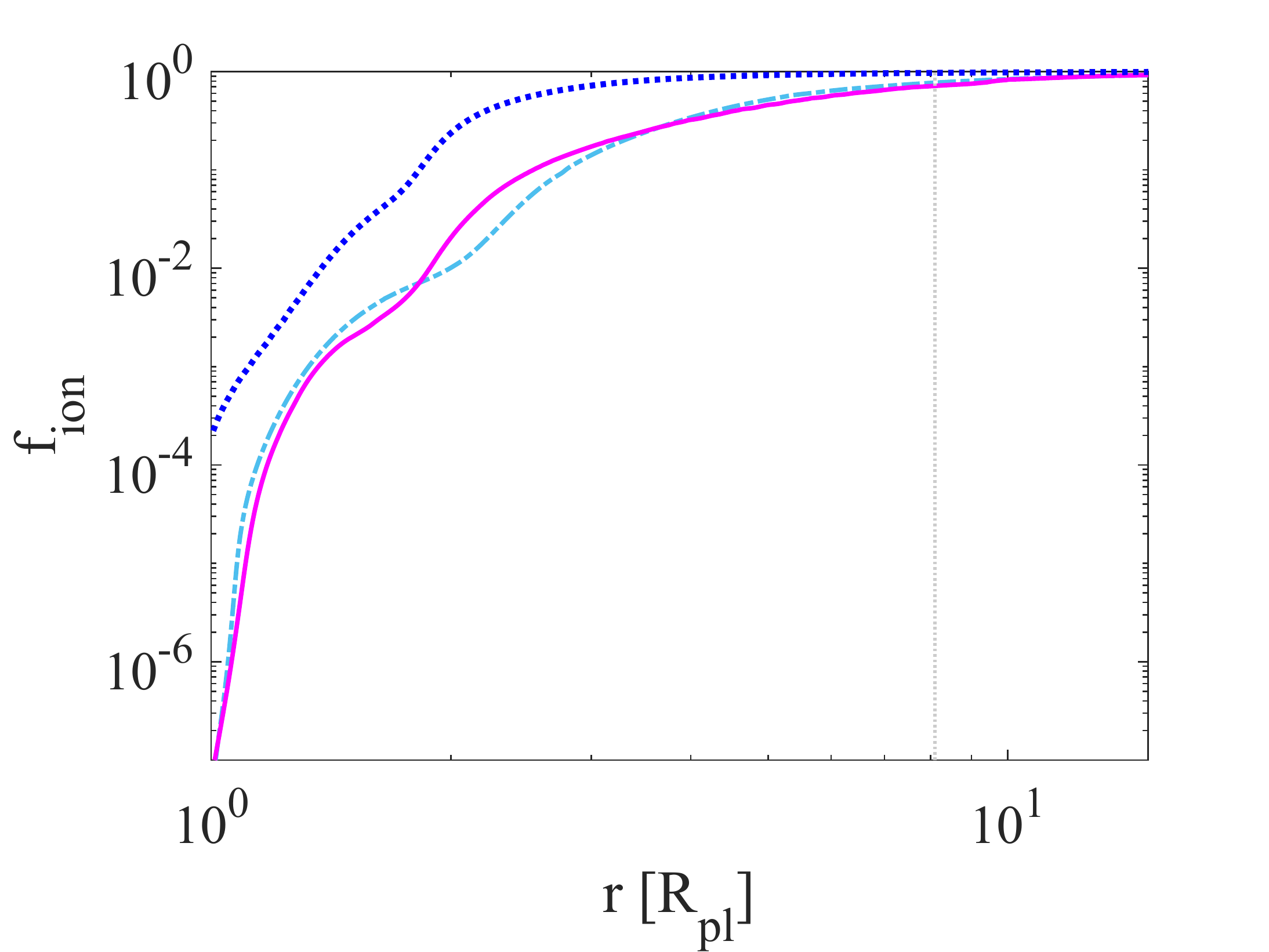}
    \caption{Ion fraction against planetary radial distance as predicted by model\,\#0 (dark-blue dotted line), model\,\#1 (light-blue dashed-dotted line), and model\,\#2 (magenta solid line) for the hot Neptune considered in Section\,\ref{sec::gridtests_HE}. The grey vertical line denotes the position of the Roche radius.}
    \label{fig::6.2b_fion_H}
\end{figure}

The comparison presented above suggests that the inclusion of Cloudy leads to a relatively small influence on the predicted outflow and specifically on the mass-loss rates. However, the changes in {the shape of the} temperature profiles (see the discussion in Section\,\ref{sec::gridtests_HE}) and ion fraction can be significant. Panels (d) and (h) of Figure\,\ref{fig::comp_hydro_H} presents a comparison of the integrated ion fractions below the Roche radius (Equation\,\ref{eq::fionint}) 
between models\,\#0, \#1, and \#2. In Equation\,(\ref{eq::fionint}), $n_{\rm ion}$ and $n_{\rm neu}$ are, respectively, the densities of ionised and neutral species. The models involving Cloudy predict lower total ion fractions compared to model\#0. The lack of ions is most significant in the lower, dense layers of the atmosphere, which give the largest input to the total ion fraction given by Equation\,\ref{eq::fionint} (see an example in Figure\,\ref{fig::6.2b_fion_H}). Therefore, for model\,\#1, the difference in $\int f_{\rm ion}$ reaches an order of magnitude, while for model\,\#2 it becomes even larger (see the grey lines in panels (d) and (h) of Figure\,\ref{fig::comp_hydro_H}). The difference is strongly correlated with planetary mass and equilibrium temperature, with the largest difference achieved for the coolest and lowest mass planets. The dependence on stellar XUV, however, appears to be weak and is most likely connected with the correlation between $F_{\rm XUV}$ and $T_{\rm eq}$ imposed by our setup.
\begin{figure}[h!]
    \centering
    \includegraphics[width=1.0\columnwidth]{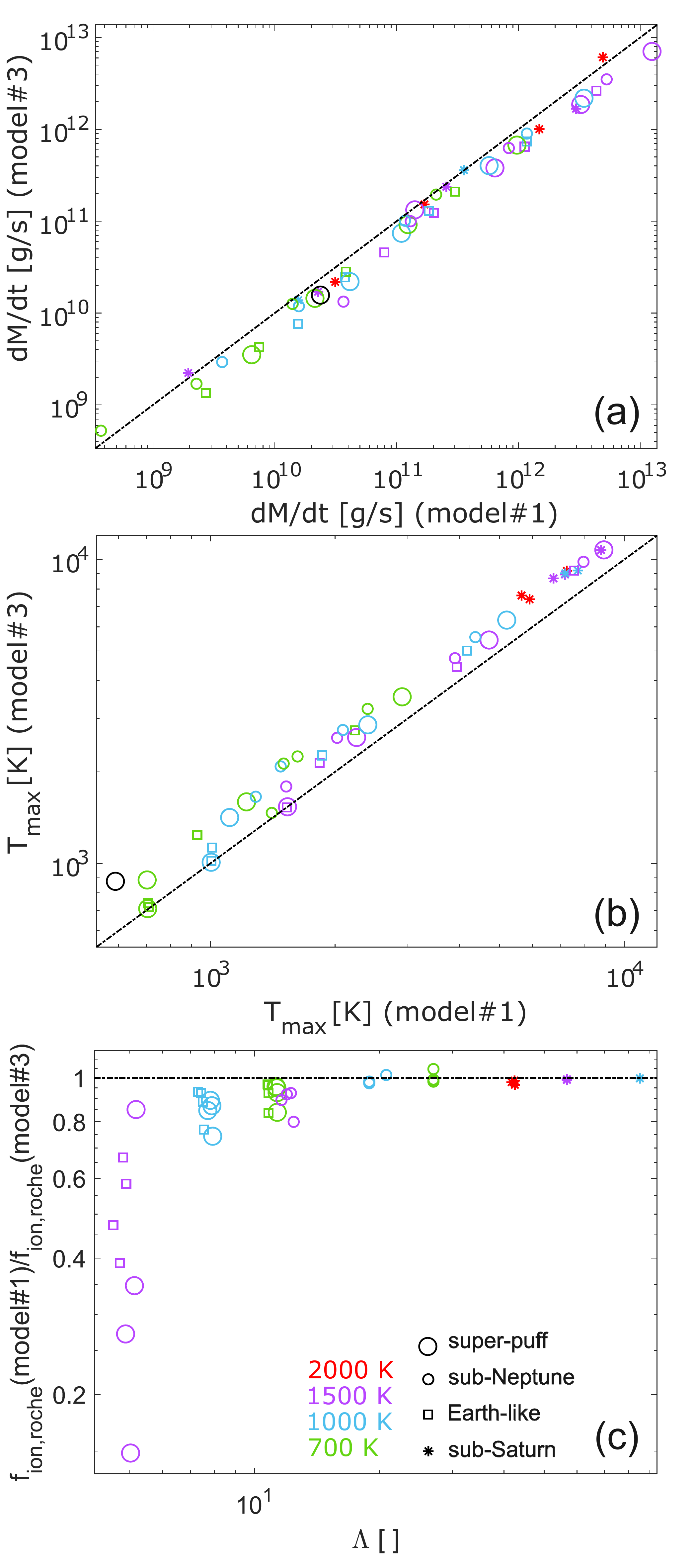}
    \caption{Comparison of the basic parameters predicted by models\,\#1 and \#3 for the set of planets presented in Section\,\ref{sec::gridtests_planets}. The colours and symbols are the same as in Figure\,\ref{fig::comp_hydro_H}, and the black lines denote the equality of parameters predicted by the two models. Panel (a): mass-loss rate. Panel (b): maximum temperature. Panel (c): ratio of the ion fraction predicted by models\,\#1 and \#3 at the Roche radius vs parameter $\Lambda$.}
    \label{fig::comp_H_He}
\end{figure}

If, instead of the total ion fraction given by Equation\,\ref{eq::fionint}, we consider the ion fraction at the position of the Roche lobe ($f_{\rm ion,roche}$), this quantity also decreases from model\#0 to model\#1, and the inclusion of wind advection in model \#2 can further decrease $f_{\rm ion,roche}$ by a factor of a few. However, the difference between $f_{\rm ion,roche}$ for model\,\#0 and \#1 (or \#2) correlates with the generalised Jeans escape parameter \citep{fossati2017} $\Lambda = (GM_{\rm pl}\mu)/(k_{\rm b}T_{\rm eq}R_{\rm pl})$ rather than with planetary mass, and it is largest for low $\Lambda$ values.
Opposite to what occurs for the total ion fraction, the differences between the ion fractions obtained from model\,\#0 and models\,\#1 and \#2 at the Roche radius, increase with increasing planetary equilibrium temperature. 

To explain the change in the dependence on $T_{\rm eq}$ between $\int f_{\rm ion}$ and $f_{\rm ion,roche}$, we consider the following.
The differences in ion fractions between the different models emerge due to differences in the treatment of the chemical network (between model\#0 and models \#1 and \#2) and to differences in atmospheric structures, i.e. due to the stronger atmospheric expansion and consequent shift of $f_{\rm ion}(r)$ upwards. The latter mechanism affects the whole atmosphere above the lowermost region dominated by the absorption of X-ray photons, while most of the chemistry occurs below 2--3\,\Rpl, where photoionisation and interactions between atmospheric particles are strongest, hence differences between the models are largest. Therefore, the differences in ion fraction at the Roche radius between models \#0 and \#1 (or \#2) are mainly driven by differences in the structural changes of the atmosphere, which increase for the hotter planets (see, e.g., the mass-loss rates and densities shown in Figure\,\ref{fig::comp_hydro_H}). At the same time, differences in the lower part of the atmosphere, and thus also the value of the total ion fraction, are mostly controlled by differences in the treatment of the chemical network by the different models.
\subsection{Effect of atmospheric composition in the upper atmospheric properties of Neptune-like planets}\label{sec::gridtests_muhe}
\subsubsection{Hydrogen atmospheres vs hydrogen-helium atmospheres}
As we discussed in Sections\,\ref{sec::code_muhe} and \ref{sec::gridtests_HE}, the differences in upper atmospheric properties predicted by the models assuming a pure hydrogen atmosphere (\#1 and \#2) and those assuming hydrogen-helium composition (\#3 and \#4) result directly from the different chemical networks involved and from differences in mean molecular weight. In Section\,\ref{sec::gridtests_HE}, we showed that for the hot sub-Neptune planet the heating is slightly larger for the hydrogen-helium atmosphere compared to the hydrogen-only atmosphere due to He{\sc i} photoionisation, while helium line cooling does not significantly contribute to the total cooling rate. This results in a less dense (but faster and hotter) outflow, and thus lower mass-loss rate, compared to the case of a pure hydrogen atmosphere (see Table\,\ref{tab::6_2B_cmodels}). Here, we compare the predictions of models\,\#1 and \#3 (not accounting for wind advection) for the set of planets presented in Section\,\ref{sec::gridtests_planets}, because the differences between models\,\#2 and \#4 (including wind advection) have similar behaviour.

Figure\,\ref{fig::comp_H_He} shows the comparison between the results obtained from models\,\#1 and \#3 for mass-loss rate, maximum temperature, and ion fraction at the Roche radius. For mass-loss rates (panel (a) of Figure\,\ref{fig::comp_H_He}), model\,\#1 predicts systematically the larger values by about a factor of two compared to model\,\#3 and this difference has a weak dependence on planetary mass and equilibrium temperature, increasing for lighter and hotter planets. The outflow density at the Roche radius behaves similarly to the mass-loss rate, but the maximum difference reaches about three times and the dependence on planetary mass is more pronounced.

As for the comparison between model\,\#0 and models\,\#1 and \#2 presented in Section\,\ref{sec::gridtests_grid}, maximum temperature and outflow velocity behave similarly. As shown by panel (b) of Figure\,\ref{fig::salz_compare}, with the inclusion of helium $T_{\rm max}$ increases up to $\sim1.5$ times and it is strongest for lower mass planets and for higher XUV fluxes. 
For $F_{\rm XUV}\gtrsim10^4$\,erg/s/cm$^2$, the differences in $T_{\rm max}$ (and $V_{\rm roche}$) between models\,\#1 and \#3 decrease steeply. %

The total ion fraction (not shown here) increases with the inclusion of helium, on average, by 10--20\%, and the difference between models\,\#1 and \#3 increases towards high XUV flux levels. Finally, the inclusion of helium modifies significantly $f_{\rm ion,roche}$ for planets with $\Lambda\lesssim7$, that is hot low-mass planets (see panel (c) in Figure\,\ref{fig::comp_H_He}). Such planets experience strong atmospheric expansion driven by their low gravity and high thermal energy, therefore the density of the outflow is large even at the position of the Roche lobe, and the ionising radiation can not penetrate very deep into the atmosphere. Therefore, most of the ionisation occurs at relatively high altitudes, where helium atoms add considerably to the ion fraction in the upper atmospheric layers. However, since the densest lower atmospheric layers remain predominantly neutral, the inclusion of helium has little impact on the total (integrated) ion fraction.
\subsubsection{Impact of the inclusion of metals}
As discussed in Section\,\ref{sec::gridtests_HE} for the hot sub-Neptune planet 4.2B, the main effect of the inclusion of metals consists in additional heating in the lowermost atmospheric layers (see the Ca{\sc ii} line heating component in Figure\,\ref{fig::6_2b_heating}). This result applies to all considered synthetic planets. As expected, the additional metal heating is weakly dependent on the stellar XUV irradiation but depends strongly on the planetary equilibrium temperature, and thus the amount of stellar UV and optical irradiation, being largest for the hottest planets. For example, for the sub-Neptune planet under moderate XUV irradiation (model planets 4.2B, 4.3B, and 4.4B in Table\,\ref{tab::gridplanets}), the maximum value of metal heating (see ${\rm Ca_{\sc ii}}$ in Figure\,\ref{fig::6_2b_heating}) increases by $\sim4.5$ times in the 1000--1500\,K $T_{\rm eq}$ range, and by $\sim2.5$ times in the 700--1000\,K $T_{\rm eq}$ range, while among these planets $F_{\rm XUV}$ varies by about 20 times. At the same time, for the hottest case of $T_{\rm eq}$\,=\,1500\,K, the maximum value of metal heating increases just about 1.5 times from the lowest to the highest XUV irradiation cases, where the difference in $F_{\rm XUV}$ spans about three orders of magnitude (or about two orders of magnitude for the EUV flux).

Therefore, at high XUV irradiation, which is typical of young planets at the beginning of their evolution, metal line heating is fully overtaken by hydrogen and helium photoionisation and does not impact the outflow parameters. When $F_{\rm XUV}$ drops below $\sim10^4$\,erg\,s$^{-1}$\,cm$^{-2}$, the peak values of metal line heating and of hydrogen and helium photoionisation heating become comparable for hot planets ($T_{\rm eq}\gtrsim1000$\,K). However, metal heating impacts mostly a very narrow region in the denser part of the atmosphere, and hydrogen and helium photoionisation occurs in 10--50 times wider atmospheric region. This limits the impact of metal heating on the atmospheric parameters, except for hot planets at low XUV irradiation (see the ``C'' model planets in Table\,\ref{tab::gridplanets}).

\begin{figure}[h!]
    \centering
    \includegraphics[width=0.97\columnwidth]{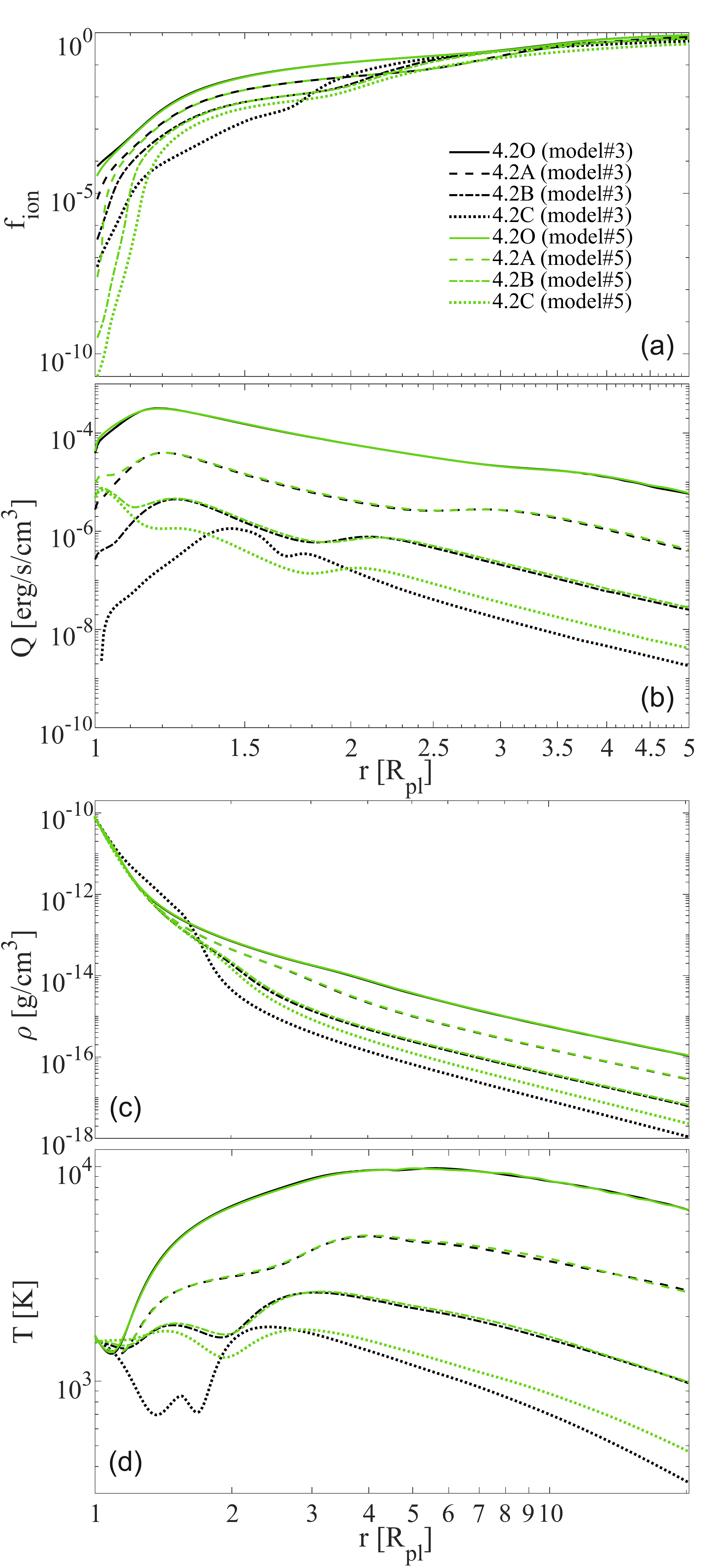}
    \caption{Atmospheric profiles for ion fraction (panel (a)), volume heating rate (panel (b)), mass density (panel (c)), and temperature (panel (d)) predicted by model\,\#3 (hydrogen-helium atmosphere, black lines) and model\,\#5 (solar abundances, green lines) for the hot sub-Neptune planet considered in Section\,\ref{sec::gridtests_HE} under different levels of XUV irradiation. Different line styles correspond to different irradiation levels: $F_{\rm XUV} = 1.5\times10^6$\,erg\,s$^{-1}$\,cm$^{-2}$ (planet 4.2O in Table\,\ref{tab::gridplanets}; solid lines), $F_{\rm XUV} = 1.8\times10^5$\,erg\,s$^{-1}$\,cm$^{-2}$ (4.2A; dashed lines), $F_{\rm XUV} = 2.7\times10^4$\,erg\,s$^{-1}$\,cm$^{-2}$ (4.2B; dashed-dotted lines), and $F_{\rm XUV} = 4.64\times10^3$\,erg\,s$^{-1}$\,cm$^{-2}$ (4.2C; dotted lines).}
    \label{fig::6.2_metals-XUV}
\end{figure}

As an example, we compare in Figure\,\ref{fig::6.2_metals-XUV} the atmospheric profiles for the hot sub-Neptune planet under different levels of XUV irradiation, i.e. model planets 4.2O--4.2C in Table\,\ref{tab::gridplanets}, predicted by model\,\#3 (black lines) and model\,\#5 (green lines). For the planet with the lowest level of $F_{\rm XUV}$ irradiation (dotted lines), the volume heating rate below $\sim1.1$\,\Rpl\ predicted by model\,\#5 reaches values similar to that given by the model run for an order of magnitude higher XUV flux (dashed-dotted lines). Therefore, for this planet, differently from the more irradiated planets, additional heating leads to more effective atmospheric expansion and denser outflow (panel (c)). The temperature profile changes drastically below $\sim2$\,\Rpl, and the maximum difference between model\,\#5, which has a comparable secondary maximum, and model\,\#3 reaches almost 1000\,K at $\sim1.45$\,\Rpl\ (compare the green and black dotted lines in panel (d) of Figure\,\ref{fig::6.2_metals-XUV}). 

Finally, the increase in heating between models\,\#3 and \#5 is accompanied by a decrease in ion fraction below $\sim1.1$\,\Rpl. Although the ionisation terms remain at a similar level in the two models (see Figure\,\ref{fig::6_2b_heating}), the additional heating leads to the upward shift of the ionisation profile. {However, as noted above, at relatively moderate planets considered in the present study, the effects discussed above can be reduced as a result of condensation. Therefore, metal heating is expected to be more prominent for hotter planets (or subject to strong XUV heating) with hydrogen-dominated atmospheres, i.e. hot Jupiters \citep{fossati2021}.}
\section{Impact of stellar spectra}\label{sec::disscussion_spectra}
In this section, we address the impact of stellar spectra on the atmospheric parameters predicted by our models. Until now, throughout this study, we employed the solar spectrum scaled to different values of X-ray, EUV, and bolometric flux as described in Section\,\ref{sec::code_SED}, which means that the results presented above can be to some extent biased by this choice. To test this possibility, we have considered the three sub-Neptune-like planets among those considered in Section\,\ref{sec::salz} for which the spectra of the host stars were constructed in the frame of the MUSCLES survey \citep[version 22; see ][]{musclesI,musclesII,musclesIII}, namely GJ\,1214\,b, GJ\,436\,b, and HD\,97658\,b. As the results obtained for these three planets are comparable, we focus here on the case of GJ\,1214\,b, and show the results for the other two planets in Appendix\,\ref{apx::spectra}.

Previous studies have shown that changes in the shape of the spectral energy distribution within the XUV band have a small influence on atmospheric mass-loss rates, but can affect considerably the temperature and ionisation profiles by altering the photochemical processes \citep[e.g.][]{guo_benjaffel2016,odert2020,kubyshkina2022_3d}, which is then important for the interpretation of observations. However, as we have shown in Section\,\ref{sec::gridtests_HE}, the heating efficiency depends on the photon energy, which might then lead to the bulk outflow parameters and mass-loss rate depending on the shape of the stellar SED in the XUV band.

\begin{figure}
    \centering
    \includegraphics[width=0.97\columnwidth]{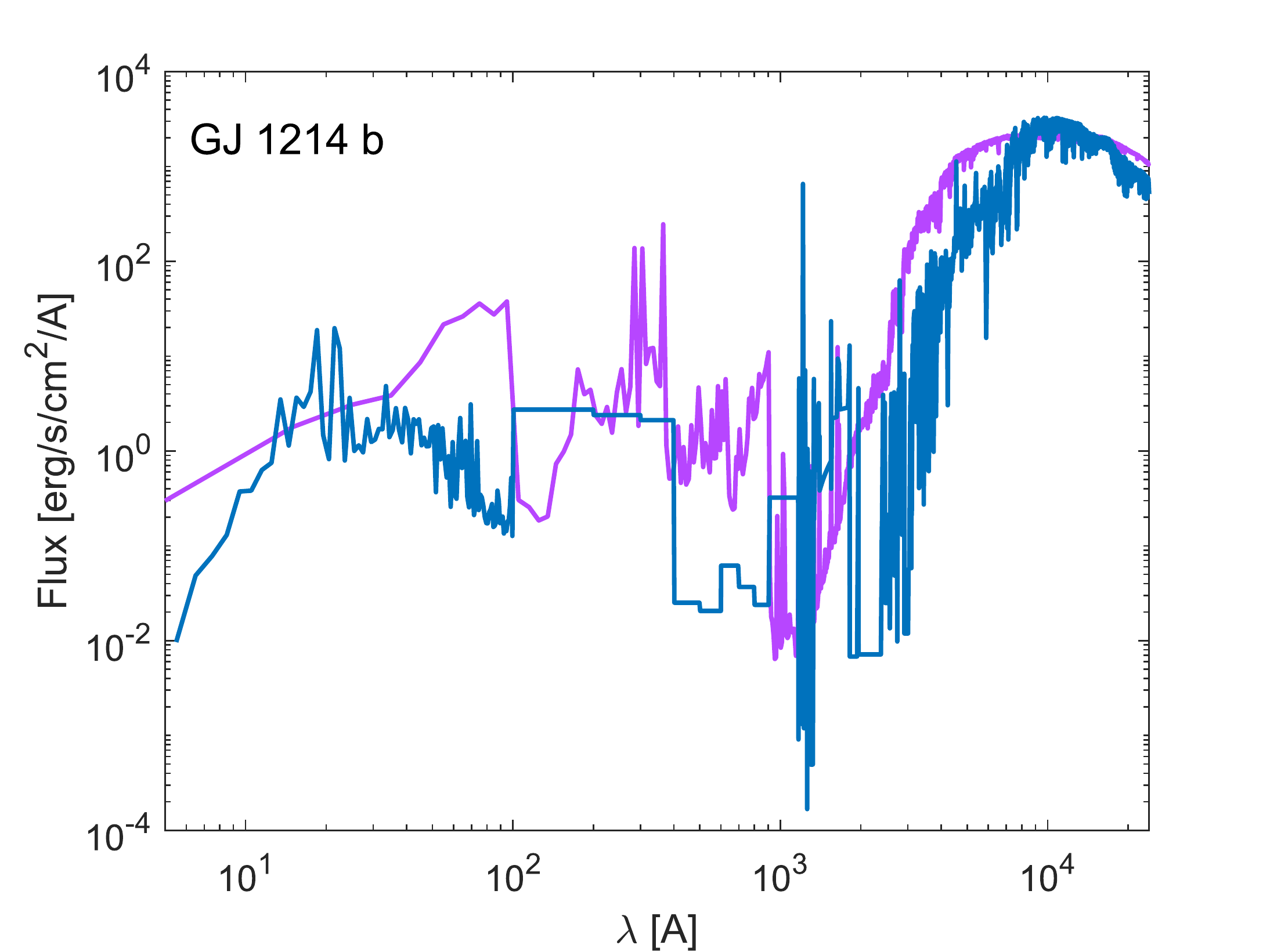}
    \caption{Spectra used to model the upper atmosphere of GJ\,1214\,b. The violet line is the scaled solar spectrum, while the blue line is the MUSCLES spectrum for GJ\,1214 scaled to the same values of $L_{\rm X}$, $L_{\rm EUV}$, and $L_{\rm bol}$ adopted by \citet{salz2016a}. The flux is scaled to the planetary orbit.}
    \label{fig::spec_gj1214}
\end{figure}

Figure\,\ref{fig::spec_gj1214} shows the comparison between the solar spectrum scaled to the luminosity of GJ\,1214 with that given by the MUSCLES survey. To reduce the comparison to the shape of the SED, we scale the MUSCLES spectrum in the same manner as the solar spectrum. The original values of $L_{\rm X} \simeq 3.6\times10^{25}$\,erg\,s$^{-1}$\,cm$^{-2}$, $L_{\rm EUV} \simeq 3.3\times10^{26}$\,erg\,s$^{-1}$\,cm$^{-2}$, and $L_{\rm bol} \simeq 1.4\times10^{31}$\,erg\,s$^{-1}$\,cm$^{-2}$ of the MUSCLES spectrum are similar to the values employed by \citet{salz2016a} ($\sim8.1\times10^{25}$\,erg\,s$^{-1}$\,cm$^{-2}$, $\sim4.1\times10^{26}$\,erg\,s$^{-1}$\,cm$^{-2}$, and $\sim1.9\times10^{31}$\,erg\,s$^{-1}$\,cm$^{-2}$ respectively). With respect to the shape of the SED, in the MUSCLES spectrum, the energy deposition in the X-ray (5--100\,\AA) and EUV (100--912\,\AA) bands is redistributed towards shorter wavelengths compared to the solar-like SED. Therefore, although the integrated flux in the respective intervals is the same in both cases, for an M-dwarf the XUV energy is, on average, transported by more energetic photons.

\begin{figure}[h!]
    \centering
    \includegraphics[width=0.89\columnwidth]{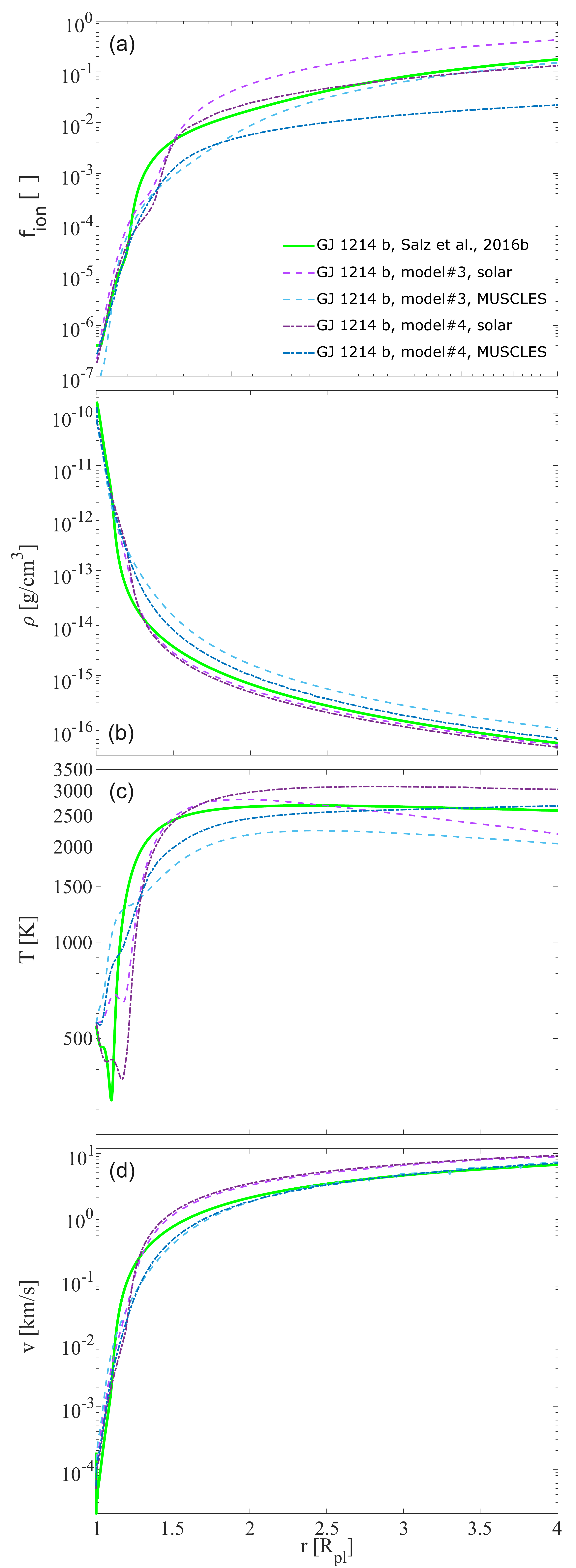}
    \caption{Atmospheric profiles of GJ\,1214\,b calculated using model\,\#3 (dashed lines) and model\,\#4 (dashed-dotted lines) and employing scaled solar (violet) and MUSCLES (blue) input stellar spectra. For reference, we show the profiles from \citet{salz2016a} in green. Panel (a): ion fraction. Panel (b): mass density. Panel (c): temperature. Panel (d): bulk velocity.}
    \label{fig::spec_gj1214_profiles}
\end{figure}

In Figure\,\ref{fig::spec_gj1214_profiles}, we show the ion fraction (a), mass density (b), temperature (c), and the bulk velocity (d) profiles for GJ\,1214\,b calculated by models\,\#3 (dashed lines) and \#4 (dashed-dotted lines) considering the solar-like spectrum (violet lines) and the MUSCLES spectrum (blue lines). The redistribution of the X-ray and EUV flux towards larger photon energies leads to a temperature increase in the lower atmospheric layers, as the higher energy photons are absorbed deeper in the atmosphere. This leads to an increase in the density by a factor of $\sim2.0$ and to a decrease in the outflow velocity by a factor of $\sim1.3$. As these two effects cancel each other in terms of the mass outflow, the atmospheric mass-loss rates computed employing the more energetic spectrum increase by $\sim1.5$ times (from $2.5\times10^{10}$\,g\,s$^{-1}$ to $4.2\times10^{10}$\,g\,s$^{-1}$ for model\,\#3 and from $2.0\times10^{10}$\,g\,s$^{-1}$ to $2.7\times10^{10}$\,g\,s$^{-1}$ for model\,\#4).

The differences in mass outflow between the results obtained considering the two different input stellar spectra are more pronounced in the case of model\,\#3, hence not accounting for the wind advection, which in practice redistributes the absorbed energy across the atmosphere. In contrast, the differences in ion fraction and temperature are larger for model\,\#4, hence including wind advection. The largest difference in the temperature profile is achieved at $\sim$1.15\,\Rpl\ and is almost a factor of two for model\,\#3 (from $\sim650$\,K to $\sim1260$\,K) and of $\sim$2.6 for model\,\#4 (from $\sim370\,K$ to $\sim950$\,K). For the ion fraction, the difference between the predictions of the models computed for different stellar spectra becomes large above $\sim1.2$\,\Rpl\ and reaches factors of 2.9--3.6 at the Roche lobe. Therefore, the total ion fraction, which is controlled mainly by the lower atmospheric layers, does not change significantly (1.3--1.4 times).

The changes in the shape of the stellar SED, and the consequent changes in atmospheric parameters are qualitatively similar for GJ\,436\,b and HD\,97658\,b (see Appendix\,\ref{apx::spectra}). However, quantitatively the difference obtained from results computed with the different input stellar spectra decreases with increasing host star mass. This is because, with increasing mass, the input spectra become closer and closer to that of the Sun that we employed as a reference. %
\section{Conclusions}\label{sec::conclusions}

We presented a new code combining hydrodynamics of planetary upper atmospheres \citep{kubyshkina2018grid} with the accurate chemistry and radiative transfer NLTE solver Cloudy. We applied this tool to model a terrestrial-like planet hosting a thin hydrogen-dominated atmosphere, a typical sub-Neptune-like planet, a low-mass super-puff, and a sub-Saturn. We aimed at testing to which extent the more accurate chemistry and radiative transfer computations provided by Cloudy affect the predictions of atmospheric outflow for different types of planets. Each of the four test planets was considered lying at different orbits, hence having different equilibrium temperatures (700--2000\,K) and under different XUV irradiation levels. The latter corresponds to what is typically expected for solar-like stars at young ages close to protoplanetary disk dispersal ($\sim1-20$\,Myr), at the end of the initial extreme phase of planetary atmospheric escape ($\sim1$\,Gyr), and towards the end of the star's main sequence lifetime ($\sim10$\,Gyr). We test also the impact of different atmospheric compositions, namely hydrogen-only atmosphere, hydrogen-helium atmosphere, and including metals in solar abundance to the hydrogen-helium atmosphere. Finally, we test the impact of accounting or not for the drag of species caused by wind advection in the Cloudy computations. In total, we considered about 50 test cases. We further compared the results with what was obtained by employing the hydrodynamic model alone. 

The approach of combining a hydrodynamic atmospheric escape model with Cloudy is not completely novel and was earlier performed by \citet[][for the hydrodynamic code PLUTO; \citealt{Mignone2007,Mignone2012}]{salz2015} and \citet[][for a Parker-wind model]{linssen2022}. Despite some differences in the numerical approach and the underlying hydrodynamic model, our code is closer to that of \citet{salz2015}. Therefore, we used the results of their model presented in \citet{salz2016a} to validate our code.

By comparing the results of the modelling, we find that the atmospheric mass-loss rates, which are of key importance for atmospheric evolution studies, do not change significantly between our basic hydrodynamic model and different model configurations involving Cloudy. The differences are largest for hot low-mass planets and for planets experiencing the highest XUV flux irradiation, but typically do not exceed a factor of two. Given that for young low and moderate mass planets, where such conditions are typical, the atmospheric mass-loss rates normally exceed $\sim10^{12}$\,g\,s$^{-1}$, such changes are not expected to significantly impact the evolution of planetary atmospheres. {Therefore, in terms of atmospheric mass-loss rates, the accurate photoionisation treatment can be more relevant for cooler planets orbiting Gyrs-old stars.}

Instead, the temperature and ionisation profiles, which are more relevant for the %
atmospheric characterisation%
, change significantly with the implementation of Cloudy in hydrodynamic modelling. The inclusion of Cloudy in the hydrodynamic computations can lead to total ion fractions of more than a factor of ten different compared to those obtained from the hydrodynamic model alone. The difference increases for planets exposed to low-to-moderate XUV fluxes, that is planets with ages older than 1\,Gyr, which comprise the majority of planets whose atmospheres are being characterised.

The specific atmospheric composition, and in particular metal line heating and cooling, appear to be most relevant for hot ($T_{\rm eq}\geq1500$\,K) planets experiencing relatively low levels of XUV irradiation ($F_{\rm XUV}\lesssim10^4$\,erg\,s$^{-1}$\,cm$^{-2}$). This supports the results obtained by \citet{fossati2021} for KELT-9\,b, which is an ultra-hot Jupiter subject to extremely low XUV irradiation. {For more moderate planets, such as those considered in this work, these effects can be decreased by condensation, which is expected to occur at altitudes similar to those of metal heating.} At very high XUV fluxes typical for young stars, all minor heating and cooling processes are suppressed by extreme photoionisation heating. The ionisation and temperature profiles in such cases are smoother compared to what obtained employing the hydrodynamic model alone. For planets in a similar mass range, the differences between models computed assuming different atmospheric compositions are more pronounced for planets with thicker atmospheres. The total ion fractions estimated for different atmospheric types do not change much. However, the distribution of the ionised species and the ion fraction in the upper layers of the atmosphere can vary considerably, specifically for hot low-mass planets ($\Lambda\lesssim7$).

The results presented here call for a detailed parameter study of the impact of considering accurate chemistry and NLTE radiative transfer computations in hydrodynamic upper atmosphere modelling. This is particularly important to more accurately interpret the wide range of already available transmission spectroscopy observations spanning from the UV to the near-infrared bands collected for hot-Jupiter to sub-Neptune-like planets.

\begin{acknowledgements}
The authors acknowledge the valuable comments from Dr. Tommi Koskinen that helped to improve significantly the quality of this manuscript.
N.V.E. acknowledges support from the Krasnoyarsk Mathematical Center financed by the Ministry of Science and Higher Education of the Russian Federation (Agreement No. 075-02-2023-912).
\end{acknowledgements}



%
%

\bibliographystyle{aa} 
\bibliography{cloudy_meets_hydro} 

\appendix

\section{List of modelled planets}\label{apx::gridplanets}

In Table\,\ref{tab::gridplanets}, we list the parameters of the model planets considered in this study, namely planetary mass, radius, equilibrium temperature, semi-major axis, and X-ray and EUV fluxes. The first column lists the reference number of the model.
\begin{table*}
	\centering
	\caption{Input parameters of the synthetic planets considered in the paper.}
	\label{tab::gridplanets}
	\begin{tabular}{lcccccr} 
		\hline
		model & \Mpl & \Rpl & \Teq & $a$ & $F_{\rm X}$ & $F_{\rm EUV}$\\
		      & [\Mer] & [\Rer] & [K] & [AU] & [${\rm erg/s/cm^2}$] & [${\rm erg/s/cm^2}$]\\
		\hline
		1.1O & 45.1 & 4.0 & 2000 & 0.0227 & $3.24\times10^6$ & $1.21\times10^6$\\
		1.1A & 45.1 & 4.0 & 2000 & 0.0227 & $3.24\times10^5$ & $2.56\times10^5$\\
		1.1B & 45.1 & 4.0 & 2000 & 0.0227 & $3.24\times10^4$ & $5.41\times10^4$\\
		1.1C & 45.1 & 4.0 & 2000 & 0.0227 & 3240.0           & $1.14\times10^4$\\
		1.2O & 45.1 & 4.0 & 1500 & 0.0404 & $1.02\times10^6$ & $3.82\times10^5$\\
		1.2A & 45.1 & 4.0 & 1500 & 0.0404 & $1.02\times10^5$ & $8.07\times10^4$\\
		1.2B & 45.1 & 4.0 & 1500 & 0.0404 & $1.02\times10^4$ & $1.71\times10^4$\\
		1.2C & 45.1 & 4.0 & 1500 & 0.0404 & 1020.0	         & 3620.0\\
		1.3O & 45.1 & 4.0 & 1000 & 0.0908 & $2.03\times10^5$ & $7.56\times10^4$\\
		1.3A & 45.1 & 4.0 & 1000 & 0.0908 & $2.03\times10^4$ & $1.60\times10^4$\\
		1.3B & 45.1 & 4.0 & 1000 & 0.0908 & 2030.0           & 3380.0\\
		1.3C & 45.1 & 4.0 & 1000 & 0.0908 & 203.0            & 717.0\\
		2.2O & 1.0 & 1.0 & 1500 & 0.0404 & $1.02\times10^6$	& $3.82\times10^5$\\
		2.2A & 1.0 & 1.0 & 1500 & 0.0404 & $1.02\times10^5$	& $8.07\times10^4$\\
		2.2B & 1.0 & 1.0 & 1500 & 0.0404 & $1.02\times10^4$	& $1.71\times10^4$\\
		2.2C & 1.0 & 1.0 & 1500 & 0.0404 & 1020.0	        & 3620.0\\
		2.3O & 1.0 & 1.0 & 1000 & 0.0908 & $2.03\times10^5$ & $7.56\times10^4$\\
		2.3A & 1.0 & 1.0 & 1000 & 0.0908 & $2.03\times10^4$ & $1.60\times10^4$\\
		2.3B & 1.0 & 1.0 & 1000 & 0.0908 & 2030.0           & 3380.0\\
		2.3C & 1.0 & 1.0 & 1000 & 0.0908 & 203.0            & 717.0\\
		2.4O & 1.0 & 1.0 & 700 & 0.1854 & $4.86\times10^4$ & $1.81\times10^4$\\
		2.4A & 1.0 & 1.0 & 700 & 0.1854 & 4860.0           & 3830.0 \\
		2.4B & 1.0 & 1.0 & 700 & 0.1854 & 486.0            & 811.0 \\
		2.4C & 1.0 & 1.0 & 700 & 0.1854 & 48.6             & 172.0 \\
		3.2O & 2.1 & 2.0 & 1500 & 0.0404 & $1.02\times10^6$ & $3.82\times10^5$\\
		3.2A & 2.1 & 2.0 & 1500 & 0.0404 & $1.02\times10^5$ & $8.07\times10^4$\\
		3.2B & 2.1 & 2.0 & 1500 & 0.0404 & $1.02\times10^4$ & $1.71\times10^4$\\
		3.2C & 2.1 & 2.0 & 1500 & 0.0404 & 1020.0	        & 3620.0\\
		3.3O & 2.1 & 2.0 & 1000 & 0.0908 & $2.03\times10^5$ & $7.56\times10^4$\\
		3.3A & 2.1 & 2.0 & 1000 & 0.0908 & $2.03\times10^4$ & $1.60\times10^4$\\
		3.3B & 2.1 & 2.0 & 1000 & 0.0908 & 2030.0           & 3380.0\\
		3.3C & 2.1 & 2.0 & 1000 & 0.0908 & 203.0            & 717.0 \\
		3.4O & 2.1 & 2.0 & 700 & 0.1854 & $4.86\times10^4$ & $1.81\times10^4$\\
		3.4A & 2.1 & 2.0 & 700 & 0.1854 & 4860.0           & 3830.0\\
		3.4B & 2.1 & 2.0 & 700 & 0.1854 & 486.0            & 811.0\\
		3.4C & 2.1 & 2.0 & 700 & 0.1854 & 48.6             & 172.0\\
		4.2O & 5.0 & 2.0 & 1500 & 0.0404 & $1.02\times10^6$ & $3.82\times10^5$\\
		4.2A & 5.0 & 2.0 & 1500 & 0.0404 & $1.02\times10^5$ & $8.07\times10^4$\\
		4.2B & 5.0 & 2.0 & 1500 & 0.0404 & $1.02\times10^4$ & $1.71\times10^4$\\
		4.2C & 5.0 & 2.0 & 1500 & 0.0404 & 1020.0	        & 3620.0\\
		4.3O & 5.0 & 2.0 & 1000 & 0.0908 & $2.03\times10^5$ & $7.56\times10^4$\\
		4.3A & 5.0 & 2.0 & 1000 & 0.0908 & $2.03\times10^4$ & $1.60\times10^4$\\
		4.3B & 5.0 & 2.0 & 1000 & 0.0908 & 2030.0           & 3380.0\\
		4.3C & 5.0 & 2.0 & 1000 & 0.0908 & 203.0            & 717.0\\
		4.4O & 5.0 & 2.0 & 700 & 0.1854 & $4.86\times10^4$ & $1.81\times10^4$\\
		4.4A & 5.0 & 2.0 & 700 & 0.1854 & 4860.0           & 3830.0\\
		4.4B & 5.0 & 2.0 & 700 & 0.1854 & 486.0            & 811.0 \\
		4.4C & 5.0 & 2.0 & 700 & 0.1854 & 48.6             & 172.0  \\
		\hline
	\end{tabular}
\end{table*}

\section{Impact of stellar spectra on the predicted atmospheric properties of GJ\,436\,b and HD\,97658\,b.}\label{apx::spectra}
\begin{figure}
    \centering
    \includegraphics[width=0.97\columnwidth]{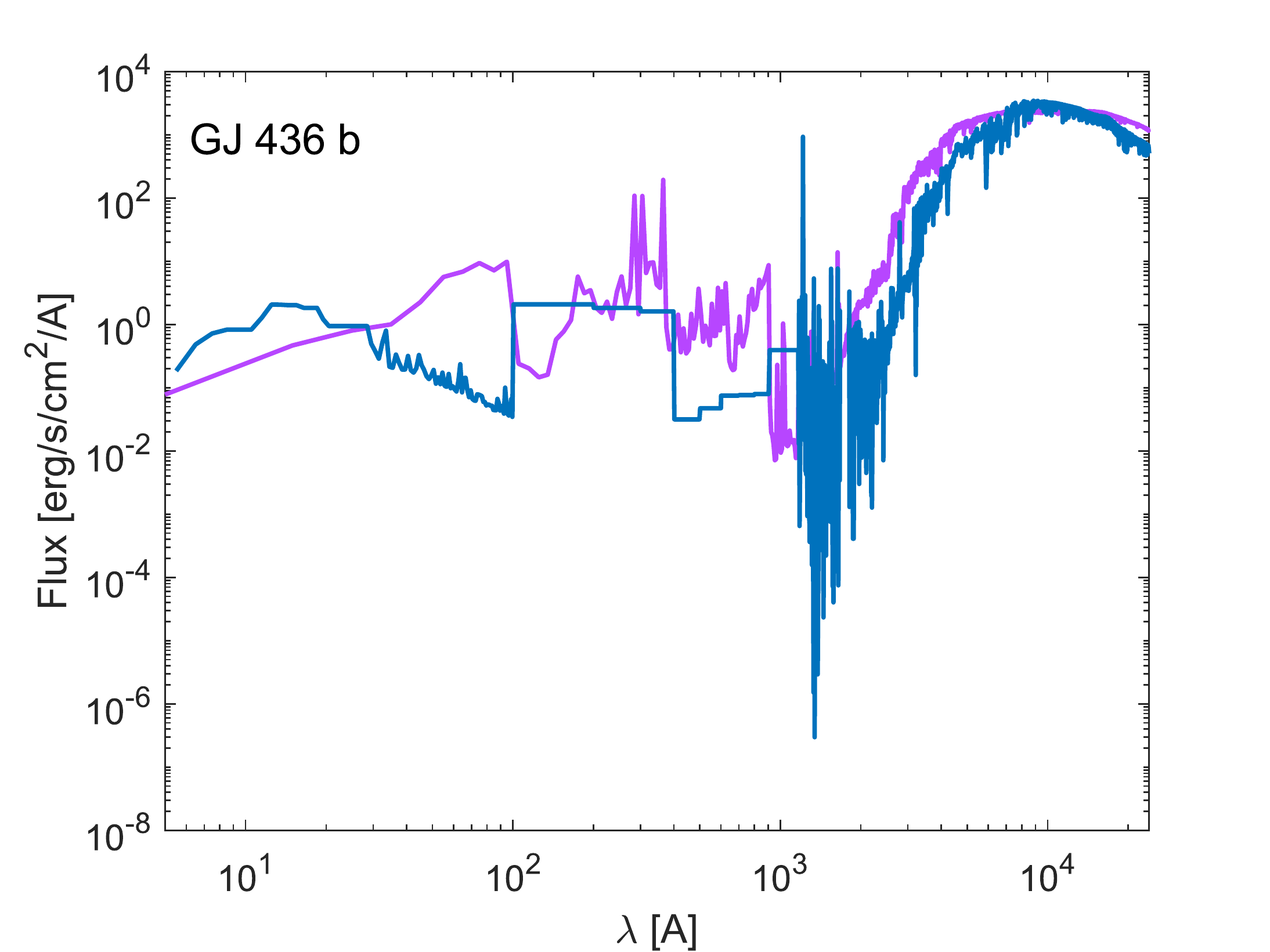}
    \caption{The same as Figure\,\ref{fig::spec_gj1214}, but for GJ\,436.}
    \label{fig::spec_gj436}
\end{figure}

\begin{figure}
    \centering
    \includegraphics[width=0.97\columnwidth]{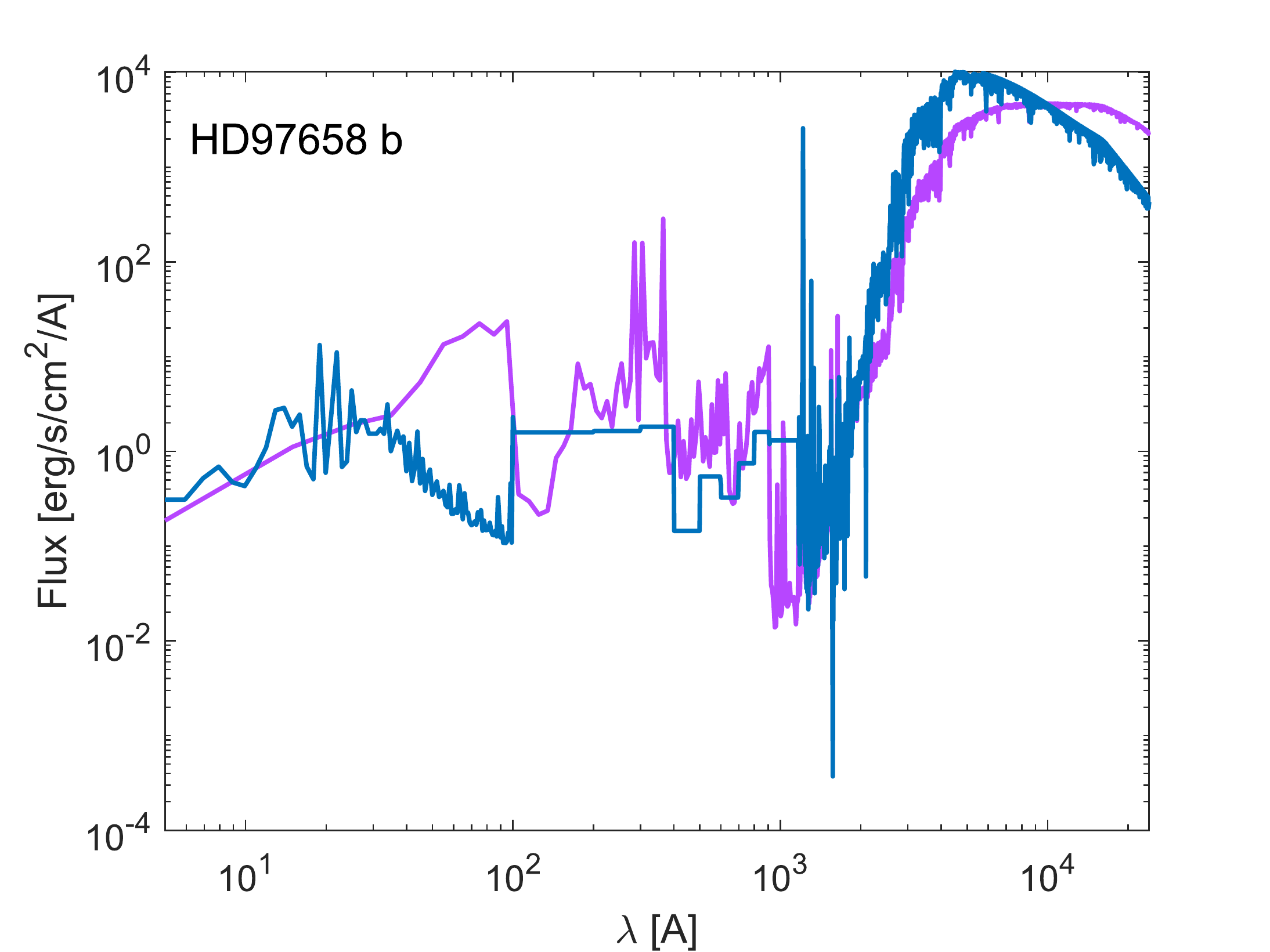}
    \caption{The same as Figure\,\ref{fig::spec_gj1214}, but for HD\,97658.}
    \label{fig::spec_hd97658}
\end{figure}

\begin{figure}
    \centering
    \includegraphics[width=0.89\columnwidth]{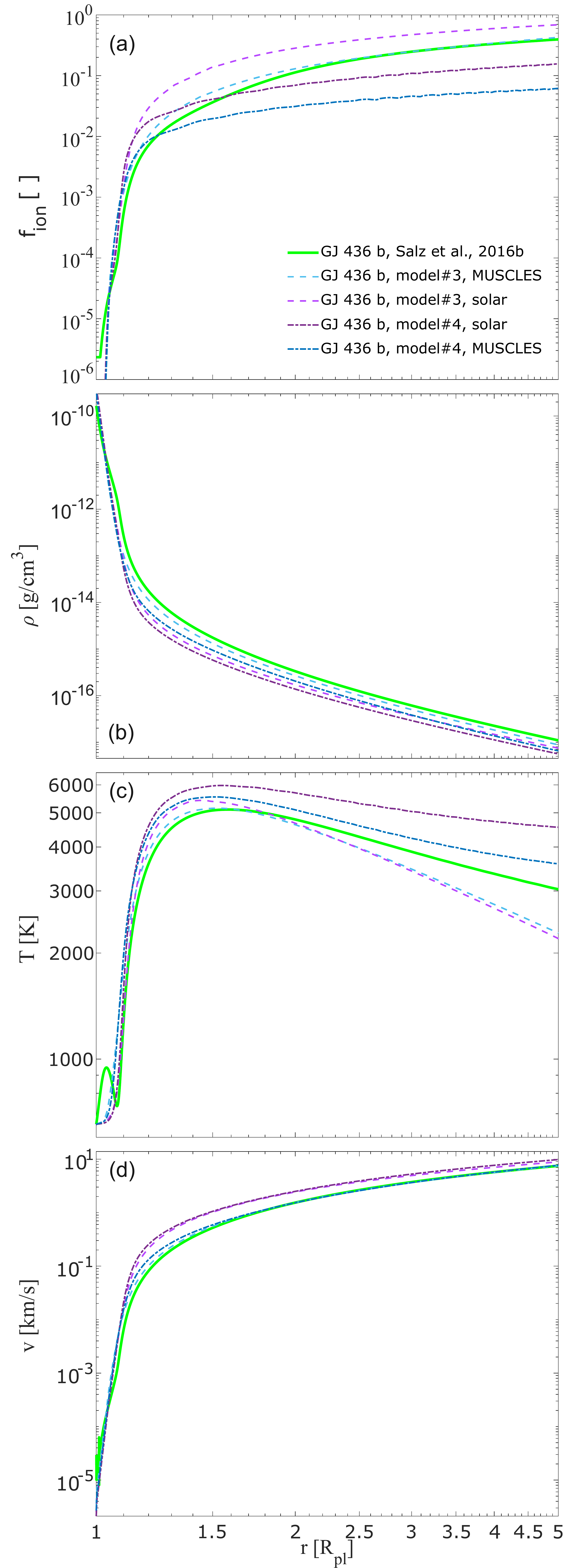}
    \caption{The same as Figure\,\ref{fig::spec_gj1214_profiles}, but for GJ\,436\,b.}
    \label{fig::spec_gj436_profiles}
\end{figure}

\begin{figure}
    \centering
    \includegraphics[width=0.89\columnwidth]{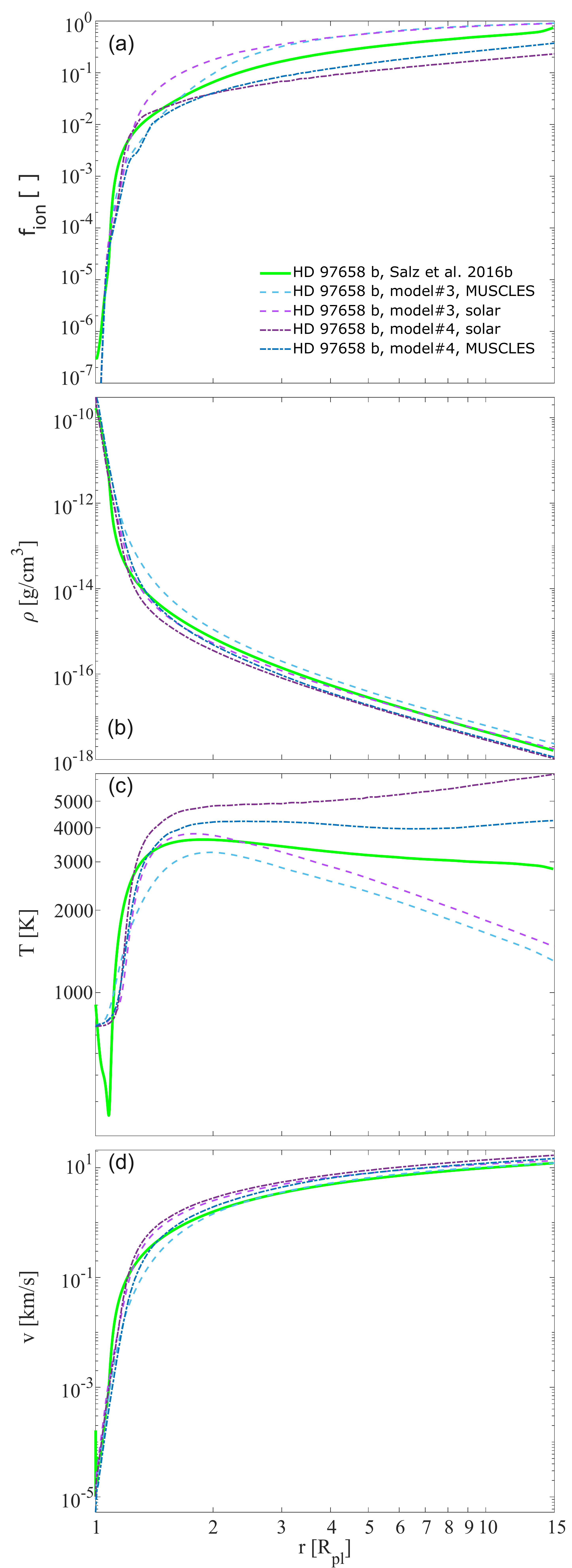}
    \caption{The same as Figure\,\ref{fig::spec_gj1214_profiles}, but for HD\,97658\,b.}
    \label{fig::spec_hd97658_profiles}
\end{figure}

Here, we present the comparison between the scaled solar spectra and the spectra constructed in a frame of MUSCLES survey for GJ\,436 (Figure\,\ref{fig::spec_gj436}) and HD\,97658 (Figure\,\ref{fig::spec_hd97658}). Figures\,\ref{fig::spec_gj436_profiles} and \ref{fig::spec_hd97658_profiles} show the atmospheric profiles obtained from models\,\#3 and \#4 employing these spectra.

\section{The changes in heating efficiency with atmospheric parameters}
\begin{figure}
    \centering
    \includegraphics[width=1\linewidth]{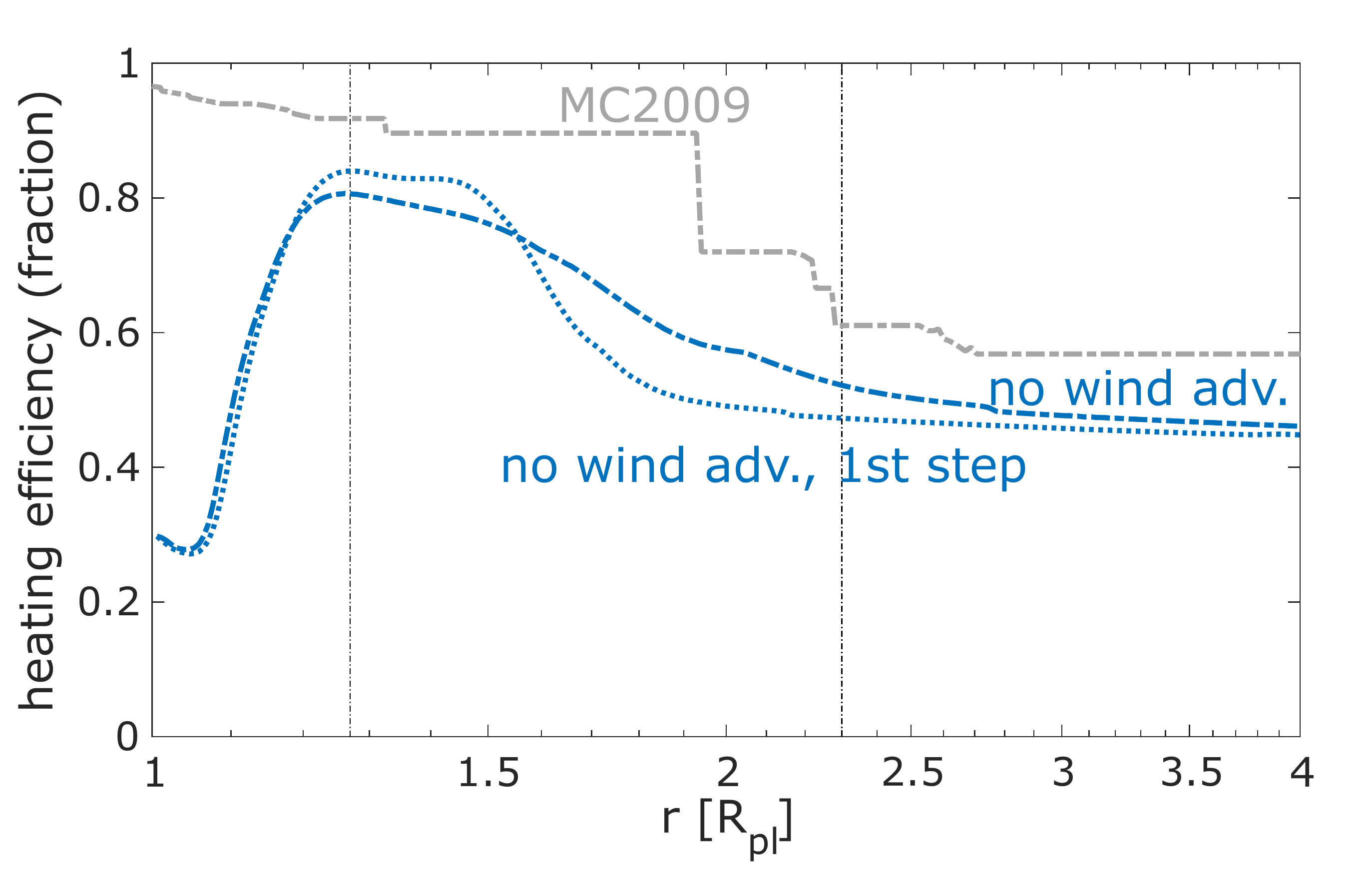}
    \caption{{Heating efficiency calculated at the first Cloudy iteration in model\,\#1 (blue dotted line) and the last Cloudy iteration of model\,\#1 (blue dashed-dotted line). The grey dashed-dotted line denotes the maximum heating efficiency calculated using the approximation from \citet{mc2009}, while the two vertical lines give the positions of the maxima of the heating function.}}
    \label{fig::HE-first-last}
\end{figure}

{Though the range of processes taken into consideration and the incoming radiation flux are being kept constant within a specific simulation, the height profile of the heating efficiency changes slightly from iteration to iteration as it follows the changes in atmospheric density and temperature. To illustrate it, we compare the heating efficiency predicted at the first iteration involving Cloudy and that of the converged solution in Figure\,\ref{fig::HE-first-last}.}

\end{document}